\documentclass[aps,pra,twocolumn, showpacs,superscriptaddress,10pt,floatfix
]{revtex4-1}

\usepackage{graphicx}
\usepackage{grffile}
\usepackage{amsthm}
\usepackage{epsfig}
\usepackage{amsmath,amssymb,amsfonts,mathtools,bbm,MnSymbol,mathrsfs,xfrac}
\usepackage{longtable}
\usepackage{tabularx}
\usepackage{multirow}
\usepackage{makecell}
\usepackage{colortbl}
\usepackage[table,svgnames,xcdraw]{xcolor}
\usepackage{array, graphicx}
\usepackage[breaklinks=true,colorlinks=true,linkcolor=blue,urlcolor=blue,citecolor=blue]{hyperref}
\usepackage{comment}
\usepackage{color}
\usepackage[makeroom]{cancel}
\usepackage{rotating}
\usepackage{tikz}
\usetikzlibrary{positioning}

\usepackage{soul}
\pdfoutput=1

\def \cc{\textcolor{red}}

\newcommand{\sect}[1]
{\medskip\noindent \textbf{#1}}
\newcommand{\ssect}[1]
{\medskip\noindent \textit{#1}}

\renewcommand{\[}{\begin{equation}}
\renewcommand{\]}{\end{equation}}

\newcommand{\ket}[1]{|#1\rangle}
\newcommand{\bra}[1]{\langle#1|}
\newcommand{\braket}[2]{\langle#1|#2\rangle}
\newcommand{\pro}[2]{|#1\rangle\langle#2|}
\newcommand{\mean}[1]{\langle#1\rangle}
\newcommand{\abs}[1]{\left|#1\right|}

\newcommand{\tr}{\mathrm{tr}}

\newcommand{\norm}[1]{\left\lvert\left\lvert#1\right\rvert\right\rvert}

\newcommand{\R}{{\hat{\rho}}}

\newcommand{\I}{{\hat{I}}}

\newcommand{\C}{{\mathcal{C}}}

\renewcommand{\P}{{\hat{P}}}
\newcommand{\Pii}{{\hat{\Pi}}}

\newcommand{\bA}{\boldsymbol{A}}
\newcommand{\bB}{\boldsymbol{B}}

\newcommand{\bx}{{\boldsymbol{x}}}
\newcommand{\by}{{\boldsymbol{y}}}

\newcommand{\bv}{{\sqrt{\boldsymbol{p}}}}

\newcommand{\ham}{{\hat{H}}}
\renewcommand{\cc}{{c_{iE}}}

\definecolor{mygray}{gray}{0.6}

\theoremstyle{definition}

\definecolor{dfcol}{cmyk}{1, 0.2108, 0.13, 0.3}
\newcommand{\df}[1]{\ifthenelse{\boolean{}}{\textcolor{dfcol}{[{\bf DF}: #1]}}{}}

\usepackage{physics}
\usepackage[page]{appendix}

\begin{document}


\title{
Measuring energy by measuring any other observable}

\author{Dominik \v{S}afr\'{a}nek}
\email{dsafranekibs@gmail.com}
\affiliation{Center for Theoretical Physics of Complex Systems, Institute for Basic Science (IBS), Daejeon - 34126, Korea}

\author{Dario Rosa}
\email{dario\_rosa@ibs.re.kr}
\affiliation{Center for Theoretical Physics of Complex Systems, Institute for Basic Science (IBS), Daejeon - 34126, Korea}
\affiliation{Basic Science Program, Korea University of Science and Technology (UST), Daejeon - 34113, Korea}

\date{\today}

\begin{abstract}
We present a method to estimate the probabilities of outcomes of a quantum observable, its mean value, and higher moments by measuring any other observable. This method is general and can be applied to any quantum system. In the case of estimating the mean energy of an isolated system, the estimate can be further improved by measuring the other observable at different times. Intuitively, this method uses interplay and correlations between the measured observable, the estimated observable, and the state of the system. We provide two bounds: one that is looser but analytically computable and one that is tighter but requires solving a non-convex optimization problem. The method can be used to estimate expectation values and related quantities such as temperature and work in setups where performing measurements in a highly entangled basis is difficult, finding use in state-of-the-art quantum simulators. As a demonstration, we show that in Heisenberg and Ising models of ten sites in the localized phase, performing two-qubit measurements excludes 97.5\% and 96.7\% of the possible range of energies, respectively, when estimating the ground state energy.
\end{abstract}

\maketitle

\section{Introduction}

Expectation values are ubiquitous in quantum physics,  characterizing different types of behaviors of quantum systems. 
They are used both as descriptive and predictive tools. To name several: mean values of generic local observables classify many-body systems according to how well they thermalize~\cite{dalessio2016from,deutsch2018eigenstate}. Vanishing total magnetization identifies a quantum phase transition~\cite{vojta2003quantum,sun2014Characterization,tian2020observation}.
The mean value of homodyne measurement~\cite{yuen1983noise,tyc2004operational,shaked2018lifting,Raffaelli2018homodyne} is evaluated in magnetic resonance imaging~\cite{noll1991homodyne} and quantum cryptography protocols~\cite{voss2009optical}, while its variance is used to prove squeezing~\cite{davidovich1996sub,Takeno2007observation} --- an essential resource for quantum sensors~\cite{lawrie2019quantum}. Variances also appear in Heisenberg's uncertainty principle~\cite{heisenberg1985anschaulichen,robertson1929uncertainty,busch2007heisenberg}. Expectation values are the object of interest in quantum field theory~\cite{wightman1956quantum,srednicki2007quantum} and in nuclear physics~\cite{BUNGE1993rooothaan,ikot2019eigensolution}.

Moments of energy are somewhat special due to their wide range of applications. The mean energy determines the thermodynamic entropy of the system~\cite{Deutsch2010thermodynamic,swendsen2015thermodynamic,santos2011entropy,safranek2021brief} and its temperature~\cite{hovhannisyan2018measuring,mukherjee2019enhanced,cenni2021thermometry}. Its change may represent heat and work~\cite{Engel2007jarzynski,rezakhani2016correlations,modak2017work,Goold2018therole,DeChiara2018ancilla,varizi2020quantum} and its difference defines a measure of extractable work called ergotropy~\cite{Allahverdyan2004a,alicki2013entanglement,safranek2022work}.
Variance in energy determines the precision in estimating both the time~\cite{paris2008quantum} and temperature~\cite{correa2015individual} parameters. Both moments, when combined, provide a tight bound on the characteristic time scale of a quantum system~\cite{Mandelstam1991uncertainty,MARGOLUS1998maximum,deffner2017quantum}. 

Given the breadth of applications, it is clear that measuring and estimating expectation values is of essential importance. This may be however challenging. For example, in quantum many-body systems, the mean energy is considerably difficult to measure, with only a few architecture-specific proposals~\cite{villa2017cavity} and experiments~\cite{jimenez2021quantum} known. This is because energy eigenstates are typically highly entangled. In quantum simulators, measuring in an entangled basis is performed by combining several elementary gates. Each gate has a fixed fidelity, and when many of such gates are combined, the fidelity diminishes making such measurements unreliable~\cite{nielsen2002quantum93,reich2013optimal,harper2017estimating,huang2019fidelity}. Additionally, experimental setups may allow measurement only of a close but not the exact observable we are interested in. This is the case, for example, in the aforementioned homodyne detection with a finite, instead of infinite, oscillator field strength~\cite{tyc2004operational,combes2022homodyne}.

In this paper, we show that performing any measurement bounds the probabilities of outcomes, the mean value, and higher moments of \emph{any other} observable.
This means that, quite unintuitively, measurements carry more information than previously known. Any observable yields \emph{some} information on \emph{any other} observable. The method uses correlations between the measured, the estimated observable, and the state of the system. It is precisely this interplay that allows us to bound the probability of outcomes of the estimated observable and, from those, its mean value and higher moments. 

These results immediately ameliorate the issue mentioned above: even in experimental systems in which we have only a limited ability to measure, we can perform the best possible measurement, and this is enough to estimate the probability distribution of outcomes and the mean value of an observable that we are truly interested in measuring.

The derived bounds are further tightened by measuring in different bases and, in the case of estimating the mean energy, by measuring at different times. After some preliminaries, we show how measurement in any basis bounds the probabilities of the system to have a certain mean energy. From this, we derive two bounds on the mean value of energy: one analytic which is easy to compute, and one tighter which leads to an optimization problem. We discuss situations in which the analytic bound becomes relatively tight. Then we describe a few differences when bounding the mean values of observables other than energy. We illustrate this method on several experimentally relevant models. Finally, we discuss the advantages and drawbacks of this method, possible applications, and future directions.

\begin{figure*}[t!]
\begin{center}
\includegraphics[width=.68\hsize]{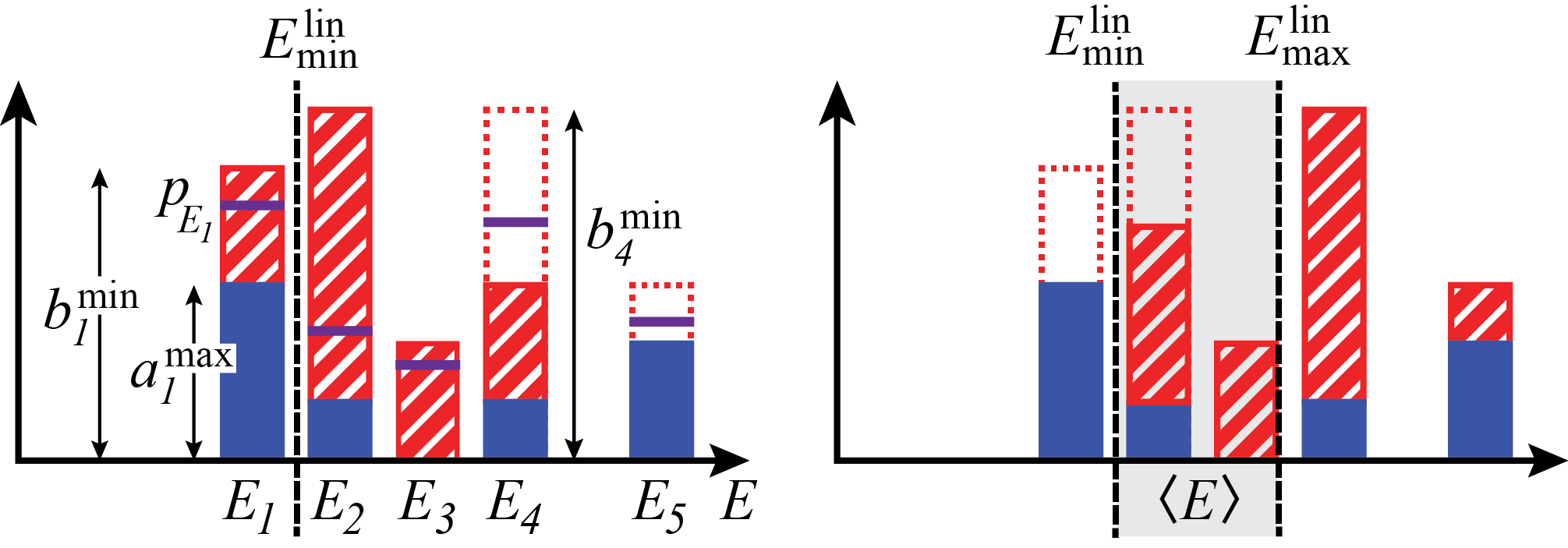}\\
\caption
{{\bf Bounds on energy probabilities and the mean energy.} Each energy probability $p_{E_l}$ (purple lines) is bounded by $a_l^{\max}$ from below (blue bar) and by $b_l^{\min}$ (red bar) from above, see Eq.~\eqref{eq:linear_inequalities}. The analytical bound on the mean energy, $E_{\min}^{\mathrm{lin}}\leq\mean{E}\leq E_{\max}^{\mathrm{lin}}$, Eq.~\eqref{eq:bounds_on_observable}, is computed as follows: imagine a bottle of probabilities with volume one. To obtain the lower bound on the mean energy (left figure), we pour the probabilities on the graph from the bottle and fill the minimal probability of each energy given by $a_l^{\max}$ (blue solid bar). We will have some probabilities in the bottle left, so, we pour the rest and top the red bars up to their maximum $b_l^{\min}$ (red striped bar), starting from the lowest to the highest energy, until we run out of probabilities in the bottle, i.e., until all the probabilities on the graph sum up to one. Taking the mean value of such distribution will yield the lower bound, see Eq.~\eqref{eq:newlower_bound}. The upper bound (right figure) is obtained by the same method but going from the highest to the lowest energy, see Eq.~\eqref{eq:newupper_bound}. The mean energy $\mean{E}$ lies somewhere in the shaded region.}
\label{Fig:bound}
\end{center}
\end{figure*}

\section{Results}
\sect{Setup.}
Consider any quantum system and measurement given by the measurement basis
$
\C=\{\ket{i}\}.
$
Label $i$ is the outcome of the measurement, and the probability of obtaining the outcome at time $t$ is
$
p_i=\abs{\braket{i}{\psi_t}}^2.
$
$\ket{\psi_t}$ is the state of the system at time $t$. If we create many realizations of the same experiment by repeating the sequence prepare-and-measure, we can build the statistics of outcomes and determine the probability distribution $\{p_i\}$. Thus, these probabilities are experimentally accessible.

Next, we consider a Hamiltonian of the system, with spectral decomposition in terms of its eigenvalues and eigenvectors being
$
\ham=\sum_E E \ketbra{E}{E}.
$
The probability of finding the system having energy $E$ is given by
$
p_E=\abs{\braket{E}{\psi_t}}^2.
$
We assume to know the Hamiltonian and its spectral decomposition. However, we presume that we are unable to measure it experimentally. In other words, we cannot perform the measurement in the energy eigenbasis. 
As we will show next, this does not stop us from estimating its probability of outcomes, and from those also the mean value of energy. Proofs for the following bounds can be found in Appendix~\ref{app:methods}.

\sect{Bounds on energy probabilities.}
The key result of this paper is that one finds the probability of the state having energy $E$ between two bounds,
\[\label{eq:time_dependent_bound_ab}
a_E\ \leq\  p_E\ \leq\  b_E,
\]
where we defined
\[\label{eq:upper_bound_pe_simple}
\begin{split}
a_E&= \max\Big\{2\sum_i c_{iE}^2-b_E,\ 0\Big\},\\
b_E&=\big(\sum_i\cc\big)^2,
\end{split}
\]
and $\cc=\sqrt{p_i}\abs{\braket{i}{E}}$, see Fig.~\ref{Fig:bound}. The last element contains both the probability of an outcome, $p_i$, and the correlation between the measured and the estimated observable, given by overlap $\abs{\braket{i}{E}}$. Thus, the above inequality connects the probability of the estimated observable to the probability of the measured observable, through the correlations between their eigenbases. We can easily derive that $b_E\leq 1$ from the Cauchy-Schwarz inequality. Thus, the upper bound on the energy probability is always non-trivial.

If the system is isolated, it evolves unitarily with the time-independent Hamiltonian $\ham$, and energy probabilities $p_E$ are also time-independent. In contrast, probabilities $p_i$ are time-dependent if the measurement basis does not commute with the Hamiltonian, and so are the bounds $a_E$ and $b_E$. This leads to an interesting observation that measuring at different times can make the bound tighter. Quantitatively, we have
\[\label{eq:linear_inequalities}
a_E^{\max}\leq p_E \leq b_E^{\min},
\]
where 
$
a_E^{\max}=\max_{t\in[0,T]}  a_E
$ and
$
b_E^{\min}=\min_{t\in[0,T]} b_E.
$

Let us discuss situations in which the bound becomes tight. First, assume that one of the measurement basis vectors is an energy eigenvector, i.e., $\ket{i}=\ket{E}$ for some $i$ and $E$. Then the bound gives $p_E=p_i$ for this specific $E$, as intuitively expected. Second, consider a situation in which we always obtain a single outcome when measuring at a specific time, i.e., $p_i=1$ for some $i$ at some time $t$. Doing this is akin to identifying the state of the system as being equal to $\ket{i}$. As a result, the bound gives an identity $p_E=\abs{\braket{i}{E}}$ for all energies $E$ so the entire energy distribution is determined exactly.

The two extreme cases just discussed suggest two possible scenarios in which the bounds perform well. The bounds are relatively tight when either the measurement basis resembles the eigenbasis of the estimated observable (in this case, the Hamiltonian), or when the state of the system comes close to one of the measurement basis vectors during its time evolution.


In addition to optimization over time, the inequalities can be further tightened by performing measurements in different bases. Defining a set of performed measurements, $\mathcal{M}=\{\C_m\}$, each measurement bounds the $p_E$ independently, so we can take
\[\label{eq:aemaxbemax_multiple}
a_E^{\max}=\max_{\C_m\in \mathcal{M},t\in[0,T]}a_E^m,\quad
b_E^{\min}=\min_{\C_m\in \mathcal{M},t\in[0,T]}b_E^m,
\]
for the bound~\eqref{eq:linear_inequalities}. $a_E^m(t)$ and $b_E^m(t)$ are defined by Eqs.~\eqref{eq:upper_bound_pe_simple} for each measurement $\C_m$. This may be helpful when there are limits on the types of measurements we can perform. For example, we can be experimentally limited to using only one- and two-qubit gates due to many-qubit gates having a low fidelity.

\sect{Bounds on collections of energy probabilities.} Additionally, we derive the following collective bounds on the energy probabilities,
\[\label{eq:time_dep_nonlinear}
\bv^T\bA_i\bv\ \leq\  p_i\  \leq\  \bv^T\bB_i\bv.
\]
The left and the right-hand sides are time-independent quadratic forms, which are defined by their elements as
\[\label{eq:defAB}
\begin{split}
 (\bA_i)_{EE'}&=(-1)^{1+\delta_{EE'}}\abs{\braket{E}{i}\!\!\braket{i}{E'}},\\ 
(\bB_i)_{EE'}&=\abs{\braket{E}{i}\!\!\braket{i}{E'}}.   
\end{split}
\]
They are applied on the vector of the square root of energy probabilities $\bv_E=\sqrt{p_E}$, which are those that we would like to estimate.

For the Hamiltonian evolution, extremizing over times of measurement yields tighter bounds
\[\label{eq:sum_inequalities_extremized}
\bv^T\bA_i\bv\ \leq\  p_i^{\min}\ \leq p_i^{\max}\ \leq\ \bv^T\bB_i\bv,
\]
where 
$p_i^{\min}=\min_{t\in [0,T]} p_i$ and  $p_i^{\max}=\max_{t\in [0,T]} p_i$.

These collective inequalities are generally non-linear in $p_E$ and neither convex nor concave. There are as many as the number of measurement outcomes. While they do not bound each energy probability separately, they provide relationships between their respective sizes. For example, one can derive quantitative statements of type: if $p_{E_1}$ is high, then $p_{E_2}$ must be low. They might require numerical methods to work with due to their non-linearity. However, they can provide a robust improvement in estimating energy in some cases. See Appendix~\ref{app:considerable_improvement} for such an example of coarse-grained energy measurements.

Similar to Eq.~\eqref{eq:aemaxbemax_multiple}, one can employ measurements in different bases. These generate more conditions for probabilities $p_E$  of type~\eqref{eq:sum_inequalities_extremized}, thus making quantitative relations between them stricter.


\begin{figure*}[t!]
\begin{center}
\includegraphics[width=.8\hsize]{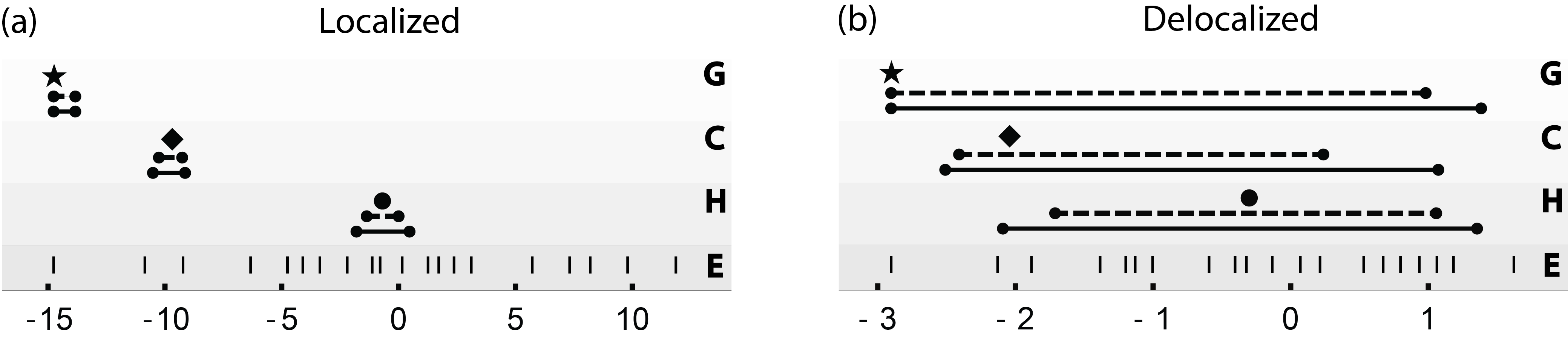}\\
\caption
{\label{tab:plots_small} {\bf Heisenberg model in the localized ($W=0.5$) and the delocalized ($W=10$) phases, 3 particles on 6 sites.} Estimating energy by measuring in the local number basis (corresponding to $k=0$ in Fig.~\ref{tab:plots_large}). The initial state is either a ground state (G), a cold state (C), Eq.~\eqref{eq:initialpurethermal}, or a hot state (H). The graphs show the true mean energy $\mean{E}$ (single symbols---star, diamond, and disc for G, C, and H, respectively), intervals of analytic bounds $[E_{\min}^{\mathrm{lin}}, E_{\max}^{\mathrm{lin}}]$ (full lines) and numerical bounds $[E_{\min},E_{\max}]$ (dashed lines) for each state. We also plot the list of energy eigenvalues at the very bottom (E). In the localized phase, energy eigenstates have a large overlap with the local number basis, making the energy estimation significantly more precise.}
\end{center}
\end{figure*}

\sect{Bounds on the mean energy.}
Given the derived bounds on the probability distribution of energy, we can bound the mean energy of the system as follows, 
\[\label{eq:bounds_on_observable}
E_{\min}^{\mathrm{lin}}\,\leq\, E_{\min}\,\leq\, \mean{E} \,\leq\, E_{\max}\,\leq\, E_{\max}^{\mathrm{lin}}.
\]
The inner bound is tighter but may be challenging to compute. The outer bound is looser, but it is analytically computable.

The inner bound is computed by optimizing the mean value of energy, $\mean{E}=\sum_E E\, p_E$, as
\[\label{eq:tight_bound}
E_{\min}=\min_{\{p_E\}\in S} \mean{E},\quad
E_{\max}=\max_{\{p_E\}\in S} \mean{E},
\]
over the set of probability distributions consistent with our observations, i.e., over the set that satisfies all the required inequalities
\[
S=\left\{\{p_E\}\bigg|\sum_E p_E=1,\ \mathrm{Eq.}~\eqref{eq:linear_inequalities},\text{ and }\mathrm{Eq.}~\eqref{eq:sum_inequalities_extremized}\right\}.
\]
The mean energy itself is a linear function. While $\sum_E p_E=1$ and Eq.~\eqref{eq:linear_inequalities} are linear constraints, Eq.~\eqref{eq:sum_inequalities_extremized} is in general non-linear.
Computing $E_{\min}$ and $E_{\max}$ is, therefore, a non-linear constrained optimization problem. These problems are considered to be computationally demanding, although they are difficult to characterize within computational complexity theory~\cite{hochbaum2007complexity}. They can be solved only approximately by various methods~\footnote{These are, for example, Nelder-Mead algorithm~\cite{LUERSEN2004globalized}, random search~\cite{price1983global}, differential evolution~\cite{storn1996usage}, machine learning methods~\cite{sivanandam2008genetic}, simulated~\cite{bertsimas1993simulated} or quantum~\cite{das2005quantum} annealing, and modified linear programming methods~\cite{powell1998direct}}. 
The time to find an exact solution typically scales exponentially with the number of variables, in our case, the dimension of the system.

However, we can solve an easier problem by including only linear constraints in the optimization, i.e., by removing the requirement for satisfying Eqs.~\eqref{eq:sum_inequalities_extremized}. This makes the bound looser but allows for solving this optimization problem analytically. The reasoning behind the following derivation is explained in Fig.~\ref{Fig:bound} and performed in detail in Appendix~\ref{app:methods}. We assume that energy eigenvalues are ordered in increasing order as $E_{j}\leq E_{j+1}$ with $E_1$ representing the ground state energy. We have
\[\label{eq:newlower_bound}
E_{\min}^{\mathrm{lin}}
=\sum_{j=1}^{N} E_j u_j,
\] 
 where probabilities $u_j\equiv u_{E_j}$ are computed recursively starting from the ground state as
\begin{align}\label{eq:recursiveprob}
u_1&=\min\left\{b_1^{\min},\ 1-\sum_{l=2}^{N}a_l^{\max}\right\},\\
u_j&=\min\left\{b_j^{\min},\ 1-\sum_{l=1}^{j-1}u_l-\sum_{l=j+1}^{N}a_l^{\max}\right\},\quad 2\leq j\leq N.\nonumber
\end{align}
(We simplified lower indices as $l\equiv E_l$, and the dimension of the system is $N$.)
Similarly, we obtain
\[\label{eq:newupper_bound}
E_{\max}^{\mathrm{lin}}=\sum_{j=1}^{N} E_j w_j,
\]
where starting from the highest energy state, we have
\begin{align}
w_{N}&=\min\left\{b_N^{\min},\ 1-\sum_{l=1}^{N-1}a_l^{\max}\right\},\\ 
w_j&=\min\left\{b_j^{\min},\ 1-\sum_{l=j+1}^{N}w_l-\sum_{l=1}^{j-1}a_l^{\max}\right\}, \quad 1\leq j\leq N-1.\nonumber
\end{align}
Bounds for the higher moments are computed by replacing eigenvalues $E$ with $E^k$ in Eq.~\eqref{eq:tight_bound}.

\sect{Tightness of the analytic bound on the mean value of energy.} 
Below Eq.~\eqref{eq:linear_inequalities}, we discussed two cases where the bound on the energy probability distribution becomes tight. Now we show that the same arguments can also be extended to discuss the tightness of the bound on the mean energy, Eq.~\eqref{eq:bounds_on_observable}.

The first case is when the state of the system comes close to one of the measurement basis vectors during its time evolution. If the state becomes exactly one of the basis vectors, i.e., $p_i(t)=1$ at some time $t$, we can identify the state exactly as $\ket{i}$ and so all of its properties, energy included. In this case, bounds~\eqref{eq:linear_inequalities} become tight and we obtain $\mean{E}= E_{\max}^{\mathrm{lin}}=E_{\min}^{\mathrm{lin}}$. The most informative times of measurement are those of a low value of observational entropy~\cite{vonNeumann1929translation,safranek2019a,safranek2019b,strasberg2021first,safranek2021brief,buscemi2022observational}, due to the state wandering into a small subspace of the Hilbert space~\cite{safranek2021brief,safranek2021generalized} recognizable by the measurement. This can be advantageous for energy estimation in systems exhibiting recurrences and Loschmidt echo~\cite{usaj1998gaussian,sanchez2016quantum,pastawski1995quantum,levstein2004nmr,rauer2018recurrences}, which return close to their original state after some time.

The second is when the measurement is close to the energy measurement itself. This happens, for example, in the localized phase of many-body localized systems, in which the energy eigenvectors tend to localize in small portions of the Fock space~\cite{abanin2019colloquium}. Thus, measuring local particle numbers is almost as good as measuring the energy itself. This is mathematically justified below Eq.~\eqref{eq:linear_inequalities}.

\sect{Choosing the time interval and times of measurements.}
In experimental settings, the system can be evolved only over a finite time. Within this time interval, only a finite number of times a measurement can be performed. Thus, it is useful to specify the criteria until which time $T$ the system should be evolved together with the corresponding times of measurement, for the time optimization, Eq.~\eqref{eq:aemaxbemax_multiple}, to work at its best.

We can address this heuristically given the points introduced in the previous section. Generally, the ideal number and times of measurement depend on the initial state: if the state does not evolve much, or at all, which is the case for any energy eigenstate (such as the ground state), then only a single measurement is required. Additional measurements will not yield any improvement. 

On the other hand, if a nontrivial evolution occurs, then more times of measurement might be advantageous. The rule of thumb is to measure for as long as the observational entropy related to the measurement grows until it reaches its equilibrium value. This is because, bigger dips in observational entropy give more information, while small dips do not provide as much. The same criterium could be applied to identify the times of measurement within this interval. There should be as many as to reproduce the medium-sized dips in the observational entropy evolution.

\sect{Bounding the mean values of observables other than energy.}
The derivations and results above can be repeated as they are for any observable that commutes with the Hamiltonian. In that context, $E$ would denote an eigenvalue of an observable other than energy. For observables that do not commute with the Hamiltonian, the procedure can be repeated but it must be performed only at a fixed time $t$ (extremization over time, Eqs.~\eqref{eq:linear_inequalities}
and~\eqref{eq:sum_inequalities_extremized}, is not possible). Extremization over different measurements at a fixed time, Eqs.~\eqref{eq:aemaxbemax_multiple}, can be employed. To summarize, the estimation of observables other than energy works exactly the same as estimating energy, with the only difference that optimization over time can be performed only when the observable commutes with the Hamiltonian. If the observable does not commute with the Hamiltonian, then also its expectation value changes in time, so only a specific time must be chosen but everything else proceeds identically.

\sect{Demonstration on experimentally relevant many-body systems.} 
We numerically demonstrate this method on the paradigmatic example of the one-dimensional disordered Heisenberg model~\cite{porras2004effective,pal2010many-body, luitz2015many}. Numerical experiments for other experimentally achieved models, Ising~\cite{smith2016many,jurcevic2017direct,zhang2017observation,bingham2021experimental}, XY~\cite{lanyon2017efficient,friis2018observation,brydges2019probing,maier2019environment}, and PXP models~\cite{bernien2017probing,Turner2018, su2022observation} are presented in Appendix~\ref{app:experimental}. A simple analytical example is presented in Appendix~\ref{app:simple}.

The Hamiltonian is given by 
\[
\ham = \sum_{i} \big(\hat \sigma_i^x\hat \sigma_{i+1}^x+\hat \sigma_i^y\hat \sigma_{i+1}^y+\hat \sigma_i^z\hat \sigma_{i+1}^z\big)+\sum_i h_i\hat \sigma_i^z,
\]
where $\hat{\sigma}_i^a$, $a = x, y, z$, are the Pauli operators acting at the site $i$. 
The constants $h_i$ are randomly extracted within the interval $\left[ -W, W \right]$ with $W$ being the disorder strength. We show the case $W = 0.5$ for the chaotic (delocalized) regime and $W=10$ for the localized regime~\cite{pal2010many-body, luitz2015many}. See Appendix~\ref{app:experimental} for the Bethe integrable regime $W=0$ \cite{Bethe1931}.

We choose a complete measurement in the local number basis,
\[
\C=\{\ket{i_1}\otimes\cdots\otimes\ket{i_L}\}_{i_1,\dots,i_L},
\]
for small systems simulations. There, $i_j=0,1$ for all $j$ and $L$ is the length of the chain (such that the dimension of the Hilbert space is $N = 2^L$). For example, for the chain of $L=3$ sites, the measurement basis is
\[
\C=\{\ket{000},\ket{001},\ket{010},\ket{011},\ket{100},\ket{101},\ket{110},\ket{111}\}.
\]
This is an example of a one-local measurement, meaning that the measurement basis does not consist of states entangled between two or more sites. For large system simulations, we also add optimized $k$-local measurements. $k$-local measurements are those that project onto states that are allowed to be entangled between $k$ neighboring sites. An example of a two-local measurement on the chain of $L=4$ sites is the measurement in the local Bell basis,
\[
\begin{split}
\C&=
\{\ket{\Phi^+}\!\ket{\Phi^+},\ket{\Phi^+}\!\ket{\Phi^-},
\ket{\Phi^+}\!\ket{\Psi^+},\ket{\Phi^+}\!\ket{\Psi^-},\dots\},
\end{split}
\]
where $\ket{\Phi^\pm}=(\ket{00}\pm\ket{11})/\sqrt{2}$ and $\ket{\Psi^\pm}=(\ket{01}\pm\ket{10})/\sqrt{2}$.

\begin{figure*}[t!]
\begin{center}
\includegraphics[width=1\hsize]{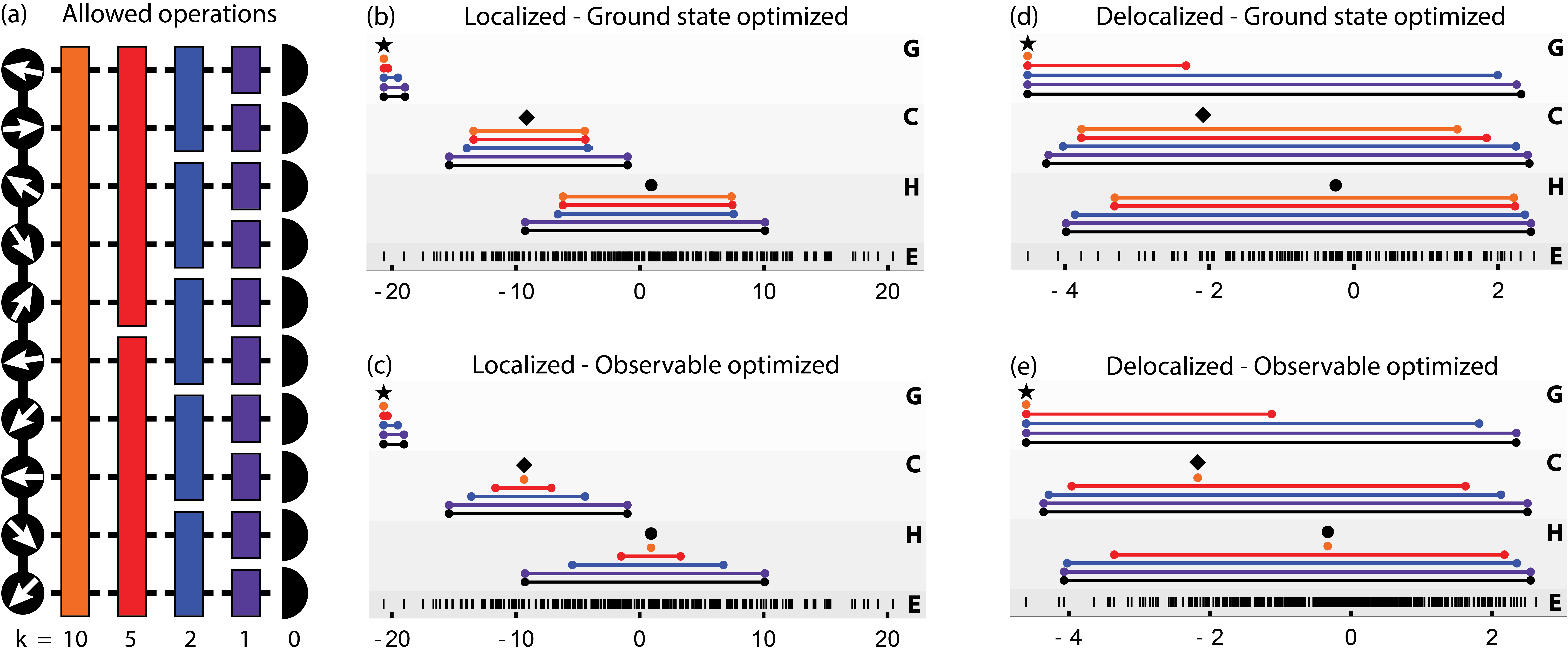}\\
\caption
{\label{tab:plots_large} {\bf Heisenberg model in the localized ($W=0.5$) and the delocalized ($W=10$) phases, 5 particles on 10 sites.} Estimating energy with progressively added optimized $k$-local measurements, with the same initial states as in Fig.~\ref{tab:plots_small} (G,C,H) and energy eigenvalues plotted at the very bottom (E). In large systems, we can no longer employ numerical methods to improve the analytic bound. Instead, we employ different measurements. (a) Sketch of allowed operations. $k=0$ (black) corresponds to measuring in the local number basis. $k=1$ (purple) allows for using single-site unitary operators before measuring in the local number basis, which leads to a general single-site measurement, $k=2$ (blue) allows using two-site local operators, etc. (b) and (d): In the ground state optimized method, the measurement basis is given by the eigenbasis of the $k$-local reduced state of the ground state, see Eqs.~\eqref{eq:klocal_ground_state} and \eqref{eq:klocal_ground_state_basis}. Intervals of analytic bounds $[E_{\min}^{\mathrm{lin (k)}}, E_{\max}^{\mathrm{lin (k)}}]$ are labeled with the colors matching the allowed operations. (c) and (e): In the observable optimized method, the experimenter measures $k$-local blocks of the Hamiltonian.  The observable-optimized method achieves perfect energy estimation for all initial states ($k=10$ corresponds to measuring the Hamiltonian itself). In contrast, the ground state-optimized achieves that only for the ground state ($k=10$ corresponds to measuring in a basis that contains the ground state). Using two-qubit measurements ($k=2$) in the localized phase, the bound excludes 97.5\% of the possible range of energies when estimating the ground state energy. See Appendix~\ref{app:methods} for the descriptions of the ground state optimized and observable optimized methods. See Appendix~\ref{app:experimental} for details and specifically Tables~\ref{tab:plots_app} and~\ref{tab:plots_app2} for additional simulations of experimentally achievable XY, PXP, and Ising models.}
\end{center}
\end{figure*}

We consider three types of initial states: First the ground state (G). The second is a ``pure thermal'' state (C),
\[\label{eq:initialpurethermal}
\ket{\psi_\beta}=\frac{1}{\mathcal{N}}\sum_E e^{-\beta E/2} \ket{E},
\]
where $\mathcal{N}$ is the normalization factor, and the inverse temperature is chosen as $\beta=6/(E_N-E_1)$.
We take this choice of $\beta$ to imitate a cold state. Third, we randomly choose a pure state from the Hilbert space with the Haar measure (H), imitating an infinite temperature state.

To compute the bounds, we evolve them with the Hamiltonian for the total time of $T=160$.

In Figure~\ref{tab:plots_small}, we show estimates of energy in small systems, taking three particles on the chain of $L=6$ sites. The Heisenberg Hamiltonian is particle conserving, so the initial state explores only a subspace of the full Hilbert space.  We analytically solve for the looser bound using Eqs.~\eqref{eq:newlower_bound} and \eqref{eq:newupper_bound}. This solution serves as a starting point for the COBYLA optimization algorithm~\cite{powell1994direct,powell1998direct} to compute the tighter bound, Eq.~\eqref{eq:tight_bound}. In Figure~\ref{tab:plots_large}, we plot estimates of energy in a large system, 5 particles on $L=10$ sites, in which computing the numeric bound is prohibitively difficult. Instead, we add more measurements and calculate the corresponding analytical bounds in the increasing degree of non-locality.

\sect{Generalization to mixed states and POVMs.}
Most general quantum measurements are represented by the positive operator-valued measure (POVM), $\C=\{\hat{\Pi}_i\}_i$, satisfying the completeness relation $\sum_i\hat{\Pi}_i=\I$. $\hat{\Pi}_i$ is a positive semi-definite operator called a POVM element. For a density matrix $\R$ representing the state of the system, the probability of obtaining measurement outcome $i$ is given by $p_i=\tr[\Pii_i\R]$.

POVM elements admit a spectral decomposition $\hat{\Pi}_i=\sum_{k} \gamma_i^k\pro{i^k}{i^k}$.
There, $0< \gamma_i^k\leq 1$ and $\ket{i^k}$ are orthogonal to each other for different $k$'s. We define its ``volume'' as $V_i=\tr[\Pii_i]$. We further define $x_{i}^E=\min_k\gamma_i^k\abs{\braket{i^k}{E}}^2$, $y_{i}^E=\max_k\abs{\braket{i^k}{E}}^2$, and $\gamma_i=\min_k\gamma_i^k$. Note that these extrema are taken only over $k$ for which $\gamma_i^k$ is positive, i.e., non-zero.

The results of this paper generalize to mixed states and general measurements by taking
\begin{align}
a_E&= \max\left\{\sum_ip_i\big(x_{i}^E+\gamma_iy_i^E\big)-\left(\sum_i\sqrt{p_i y_{i}^EV_i}\right)^2\!\!\!,0\right\},
\nonumber\\
b_E&=\bigg(\sum_i\sqrt{p_i\bra{E}\Pii_i\ket{E}}\bigg)^2,\label{eq:upper_bound_pe_POVM}
\end{align}
in Eq.~\eqref{eq:time_dependent_bound_ab}, and by taking
\[\label{eq:defAB2}
\begin{split}
 (\bA_i)_{EE'}&=(-1)^{1+\delta_{EE'}}\abs{\bra{E}\Pii_i\ket{E'}},\\
(\bB_i)_{EE'}&=\abs{\bra{E}\Pii_i\ket{E'}},
\end{split}
\]
in Eq.~\eqref{eq:defAB}. See Appendix~\ref{app:methods} for the proofs. Results that come after do not depend on the specific form of the bounds. Thus these proceed identically. For a complete projective measurement, we have $\Pii_i=\ketbra{i}{i}$, which implies $x_i^E=y_i^E=\abs{\braket{i}{E}}^2$ and $\gamma_i=V_i=1$, from which the initial bounds easily follow. 

\section{Discussion and Conclusion}

Quantum measurements provide more information than one would initially think. We developed a method that allows us to measure one observable and predict bounds on the distribution of outcomes and expectation values of every other observable. In this method, it is assumed that we have enough copies of the initial state so we can determine the entire probability distribution of outcomes of the measured observable.
The method works well either when the measured and the estimated observables resemble each other or when the system state is close to one of the measurement basis states. In those cases, the bounds  will be very tight. On the other hand, if the measurement cannot distinguish between two eigenstates with very different eigenvalues, and the system state has considerable overlap with one of them, the method naturally cannot give a good estimate. However, this can be overcome by combining measurements in different bases. Additionally, when estimating conserved quantities, better estimates are obtained by measuring at different times.

It is interesting to compare the presented method with the recent work of~\cite{huang2020predicting}. There, an algorithm is provided in which it is possible to approximate the mean value of an observable with high probability by applying random measurements, called \emph{classical shadows}. This idea has been extended in subsequent literature both theoretically~\cite{zhao2021,Hadfield2022,bu2022,sack2022,ippoliti2023,hu2022hamiltonian,hu2023,gresch2023,akhtar2023scalableflexible,seif2023shadow}, and experimentally~\cite{zhang2021,struchalin2021}. 

In classical shadows literature, it is assumed that performing any type of measurement is possible. These measurements are sampled randomly from a tomographically complete set, meaning that with this set of measurements, quantum tomography is possible. The goal is to estimate the expectation value of an observable and achieve an error lower than $\epsilon$, with as few measurements as possible. In other words, it is assumed that only a limited number of copies of the state are available to be measured.

In contrast, in the method presented here, it is assumed that only a single type of measurement can be performed. This measurement has been chosen from a limited set of measurements that are experimentally available. At the same time, it is assumed that infinitely many copies of the initial state are available. Thus, we can perform as many repetitions of the same measurement as necessary to fully specify its probability distribution. 

In our method, since we do not sample from a tomographically complete set, we always have a \emph{finite} error in estimating the expectation value. This error comes from the misalignment of the chosen measurement basis and the estimated observable and the eigenbasis of the density matrix. 

Thus, while attempting to address a very similar goal, the two approaches are different in their assumptions and outcomes. Our method shines in exactly those situations in which not every measurement can be performed.
This is motivated by the experimental capabilities of current state-of-the-art quantum simulators~, which allow for the application of only one and possibly two-qubit gates. For this reason, we focused on local measurements.  Of course, the method will work better with the improvement of the experimental capabilities. Its main strength, though, is to be able to give a prediction of the mean value of observables together with its error even in cases when the experimental capabilities are very low and other methods cannot be used.

We argued for using this method for estimating moments of energy, which have a wide range of applications while being difficult to measure directly in many-body systems. For instance, using this method, one can bound the characteristic timescale through the Mandelstam-Tamm and Margolus-Levitin bounds~\cite{Mandelstam1991uncertainty,MARGOLUS1998maximum,deffner2017quantum}, to estimate the amount of extractable work from an unknown source of states~\cite{safranek2022work}, or to estimate temperature. Moreover, the latest can be used to benchmark the cooling function of quantum annealers~\cite{benedetti2016estimation,pino2020mediator,hauke2020perspectives,imoto2021improving} and adiabatic quantum computers~\cite{mohseni2019error}.
This could be particularly suitable for systems with area-law entanglement scaling~\cite{eisert2010colloquium,abanin2019colloquium}, in which local measurements should be more powerful given the absence of long-range correlations in the eigenstates of such systems. We confirmed this numerically in two gapped models, which are proven to have area-law ground states~\cite{Hastings_2007}. In these, estimating the ground state energy using only local measurements works especially well. The method can be used equally well and proceeds identically for estimating observables other than energy, with the only exception that if such an observable does not commute with the Hamiltonian, then also its mean value is time-dependent, therefore, one has to pick a specific time for the analysis. (In contrast, when measuring energy, we could improve the bounds by measuring at different times, using that its mean value is conserved.)
The method can also be used to prove entanglement through an entanglement witness (operator $\hat{A}$), without measuring the witness itself: to prove entanglement, it is enough to show that the expectation value of this operator is negative~\cite{terhal2000bell,barbieri2003detection,lewenstein2000optimization}.

On a theoretical ground, this research instigates new possible paths of exploration. How to choose a measurement given some restriction, for example, on its locality or the number of elementary gates it consists of, that leads to the tightest possible bound? Is it possible to apply machine learning models to find this optimal strategy? Will the bound give an exact value when the set of measurements is tomographically complete? Can this method be modified to identify the properties of channels instead of states? 
Given that this method bounds the entire probability distribution of outcomes, is it possible to modify it to calculate estimates of functions of a state other than expectation values, such as entanglement entropy?

\sect{Acknowledgments.} We thank Felix C. Binder for the collaboration on a related project during which some of the ideas for this paper started to surface. We thank Dung Xuan Nguyen, Sungjong Woo, 
and Siranuysh Badalyan for their comments and discussions. We acknowledge the support from the Institute for Basic Science in Korea (IBS-R024-D1).

\sect{Author Contributions.} D.\v{S}. Conceptualization, theory, bounds and their proofs in particular, writing, visualization of figures, development of the ground state optimized and the observable optimized (type 1) methods, and creating a single qubit example in the Appendix. D.R. Development of the software for generating numerical experiments of Heisenberg, XY, PXP, and Ising models and producing the data, editing, and development of the observable optimized (type 2) method. Both authors contributed to the style of the paper and cross-contributed to the main roles of the other through frequent discussions.

\sect{Data and code availability.} Data and the code used to generate data for this paper are available from the corresponding authors upon reasonable request.

\section*{Appendix}

This Appendix provides methods and proofs of the bounds, several examples, and additional numerical experiments on experimentally realized many-body systems. It contains App.~\ref{app:methods}: Methods and proofs of the bounds. App.~\ref{app:considerable_improvement}: Example in which the bound on the collection of energy probabilities provides a much better estimate of the mean energy value. App.~\ref{app:quality}: Introduction of quality factors --- two measures of performance. App.~\ref{app:simple}: Simple analytic example of the mean energy estimation of a qubit. App.~\ref{app:experimental}: Estimation of the mean energy in experimentally relevant models: Heisenberg, Ising, XY, and PXP models.

\appendix

\section{Methods and proofs}\label{app:methods}

\sect{Ground state optimized and observable optimized methods} for finding appropriate $k$-local measurements. We choose a chain of length $L=10$ and two-local ($k=2$) measurements to illustrate.

Ground state optimized method: This method is inspired by the Matrix Product State ansatz~\cite{orus2014a}, and by the correspondence between observational and entanglement entropy~\cite{schindler2020quantum}.  We choose a pure state (in our case, the ground state) $\ket{\psi_0}$ for which we want to optimize. We divide the chain into $L/k$ local parts and generate the local basis as an eigenbasis of the reduced state. This corresponds to the local Schmidt basis. For example, for the first two sites denoted as subsystem $A_1$, while denoting the full system as $A$, we have
\[\label{eq:klocal_ground_state}
\R_{A_1}=\tr_{A\setminus {A_1}}\ketbra{\psi_0}{\psi_0},
\]
from which we compute eigenvectors $\{\ket{\psi_{i_1}^{A_1}}\}$. Eq.~\eqref{eq:klocal_ground_state} is what we refer to as the $k$-local reduced state of the ground state. The final, ground state-optimized two-local measurement basis is given as a product of such generated local basis,
\[\label{eq:klocal_ground_state_basis}
\big\{\ket{\psi_{i_1}^{A_1}}\otimes\ket{\psi_{i_2}^{A_2}}\otimes\ket{\psi_{i_3}^{A_3}}\otimes\ket{\psi_{i_4}^{A_4}}\otimes\ket{\psi_{i_5}^{A_5}}\big\}.
\]
This method ensures that for the full system ($k=L$), the ground state is one of the measurement basis states.

Observable optimized method: This method is a $k$-local optimization for a specific observable, in our case the Hamiltonian. The basis is generated by the eigenbasis of the Hamiltonian where interaction terms between the local parts have been taken out. For example, in the Heisenberg model, the two-local measurement basis is given as the eigenbasis of the Hamiltonian
\[
\ham_2 = \sum_{i=1,3,5,7,9} \big(\hat \sigma_i^x\hat \sigma_{i+1}^x+\hat \sigma_i^y\hat \sigma_{i+1}^y+\hat \sigma_i^z\hat \sigma_{i+1}^z\big)+\sum_{i=1}^{10} h_i\hat \sigma_i^z.
\]
This method ensures that the measurement basis for the full system ($k=L$) is the same as the eigenbasis of the observable we optimize for. See Appendix~\ref{app:experimental}, Methods for optimizing local measurements, for details.

\sect{Upper bound on energy probabilities. }
First, we prove an upper bound on the energy probability, 
\[\label{eq:pEbound_app}
p_E\leq \bigg(\sum_i\sqrt{p_i}\sqrt{\bra{E}\Pii_i\ket{E}}\bigg)^2,
\]
where the right-hand side defines $a_E$. This is the second half of Eq.~\eqref{eq:upper_bound_pe_POVM}.

\begin{proof} 
We express the spectral decomposition of the density matrix as $\R=\sum_m\lambda_m\pro{\psi_m}{\psi_m}$,
where $\lambda_m$ are its eigenvalues corresponding to eigenvectors $\ket{\psi_m}$.
We find that $p_i=\tr[\Pii_i\R]$ translates to $
p_i=\sum_{m}\lambda_m\bra{\psi_m}\Pii_i\ket{\psi_m}$. 
A series of inequalities follows:
\[
\begin{split}
p_E&=\bra{E}\R\ket{E}=\sum_m\lambda_m\abs{\braket{E}{\psi_m}}^2
=\sum_m\lambda_m\Big|\!\sum_i\bra{E}\Pii_i\ket{\psi_m}\!\Big|^2\\
&\leq \sum_m\lambda_m\bigg(\sum_i\abs{\bra{E}\Pii_i\ket{\psi_m}}\bigg)^2\\
&= \sum_m\lambda_m\bigg(\sum_i\abs{\bra{E}\sqrt{\Pii_i^\dag}\sqrt{\Pii_i}\ket{\psi_m}}\bigg)^2\\
&\leq \sum_m\lambda_m\bigg(\sum_i\sqrt{\bra{E}\Pii_i\ket{E}\bra{\psi_m}\Pii_i\ket{\psi_m}}\bigg)^2\\
&= \sum_m\lambda_m\left\|\bx^m\right\|_{\frac{1}{2}}\\
&\leq \Big\|\sum_m\lambda_m\bx^m\Big\|_{\frac{1}{2}}\\
&= \bigg(\sum_i\sqrt{\sum_{m}\lambda_m\bra{E}\Pii_i\ket{E}\bra{\psi_m}\Pii_i\ket{\psi_m}}\bigg)^2\\
&= \bigg(\sum_i\sqrt{p_i}\sqrt{\bra{E}\Pii_i\ket{E}}\bigg)^2.\\
\end{split}
\]
The first inequality is the triangle inequality, the second the Cauchy-Schwarz inequality, and the third Jensen's theorem applied on ($p=\frac{1}{2}$)-seminorm
\[
\left\|\bx\right\|_{\frac{1}{2}}=\Big(\sum_i\sqrt{x_i}\Big)^2.
\]
$\bx^m=(x_1^m,x_2^m,\dots)$ is a vector of positive entries $x_i^m=\bra{E}\Pii_i\ket{E}\bra{\psi_m}\Pii_i\ket{\psi_m}\geq 0$. In order to apply Jensen's theorem, we need first to confirm that the ($p=\frac{1}{2}$)-seminorm is a concave function. It is concave when restricted on vectors with positive entries, $\left\|~\right\|_{\frac{1}{2}}:\mathbb{R}_+^{N}\rightarrow \mathbb{R}_+$, which is indeed the case here. This follows from the reverse Minkowski inequality
\[
\norm{\bx+\by}_p\geq \norm{\bx}_p+\norm{\by}_p,
\]
(where it is assumed $x_i\geq 0$, $y_i\geq 0$), which holds for all ($p< 1$)-seminorms. Taking $0 \leq q\leq 1$, we have
\[
\begin{split}
 \norm{q\bx+(1-q)\by}_p&\geq \norm{q\bx}_p+\norm{(1-q)\by}_p\\
 &=q\norm{\bx}_p+(1-q)\norm{\by}_p,   
\end{split}
\]
so $\left\|~\right\|_{\frac{1}{2}}$ is indeed concave.
\end{proof}

\sect{Lower bound on energy probabilities. }
Next, we prove a lower bound on the energy probabilities,
\[\label{eq:lower_bound_methods}
p_E\geq\max\left\{\sum_ip_i\big(x_{i}^E+\gamma_iy_i^E\big)-\left(\sum_i\sqrt{p_i y_{i}^EV_i}\right)^2\!\!\!,0\right\}.
\]
where the right-hand side defines $b_E$. This is the first half of Eq.~\eqref{eq:upper_bound_pe_POVM}, above which $x_i^E$, $y_i^E$, and $V_i$ are defined. We drop superscripts $E$ to keep the notation cleaner, i.e., we write $x_i\equiv x_i^E$ and $y_i\equiv y_i^E$. 

\begin{proof}
It is clear that $p_E\geq 0$. To derive the first inequality, we start by expressing $p_E$ more conveniently as
\[
\begin{split}
p_E&=\bra{E}\R\ket{E}=\sum_m\lambda_m\abs{\sum_i\bra{E}\Pii_i\ket{\psi_m}}^2\\
&=\sum_m\lambda_m\abs{\sum_{i,k}\gamma_i^k\braket{E}{i^k}\braket{i^k}{\psi_m}}^2\\
&=\sum_{m,i,k}\lambda_m\abs{\gamma_i^k\braket{E}{i^k}\braket{i^k}{\psi_m}}^2\\
&+\sum_{m,i\neq i',k,k'}\lambda_m\gamma_i^k\gamma_{i'}^{k'}\braket{i'^{k'}}{E}\braket{E}{i^k}\braket{i^k}{\psi_m}\braket{\psi_m}{i'^{k'}}\\
&+\sum_{m,i,k\neq k'}\lambda_m\gamma_i^k\gamma_{i}^{k'}\braket{i^{k'}}{E}\braket{E}{i^k}\braket{i^k}{\psi_m}\braket{\psi_m}{i^{k'}}\\
&\geq \!\!\!\sum_{m,i,k}\!\!\lambda_m\abs{\gamma_i^k\braket{E}{i^k}\braket{i^k}{\psi_m}}^2\!\!+\!\!\!\sum_{m,i,k}\!\!\!\lambda_m\abs{\gamma_i^k\braket{i^k}{\psi_m}}^2\! y_i\!-\!w\\
&\geq \sum_{m,i,k}\lambda_m\gamma_i^k\abs{\braket{i^k}{\psi_m}}^2(x_i+\gamma_iy_i)-w\\
&=\sum_i p_i(x_i+\gamma_iy_i)-w.
\end{split}
\]
The first inequality follows from the definition of absolute value and the second from definitions of $x_i$ and $\gamma_i$. We defined
\[
\begin{split}
w&=\sum_{m,i\neq i',k,k'}\!\!\lambda_m\gamma_i^k\gamma_{i'}^{k'}\abs{\braket{i'^{k'}}{E}\braket{E}{i^k}\braket{i^k}{\psi_m}\braket{\psi_m}{i'^{k'}}}\\
&+\sum_{m,i,k\neq k'}\!\!\lambda_m\gamma_i^k\gamma_{i}^{k'}\abs{\braket{i^{k'}}{E}\braket{E}{i^k}\braket{i^k}{\psi_m}\braket{\psi_m}{i^{k'}}}\\
&+\sum_{m,i,k}\lambda_m\abs{\gamma_i^k\braket{i^k}{\psi_m}}^2 y_i\\
&\leq \sum_{m,i,i', k, k'}\lambda_m \gamma_i^k\gamma_{i'}^{k'}\abs{\braket{i^k}{\psi_m}\braket{\psi_m}{i'^{k'}}}\sqrt{y_{i}y_{i'}}\\
&=\sum_{i,i'}\!\sqrt{y_{i}y_{i'}}\sum_m\!\lambda_m\Big(\sum_{k}\! \gamma_i^k\abs{\braket{i^k}{\psi_m}}\Big)\Big(\sum_{k'}\!\gamma_{i'}^{k'}\abs{\braket{i'^{k'}}{\psi_m}}\Big)\\
&\leq \sum_{i,i'}\!\sqrt{y_{i}y_{i'}}\sum_m\!\lambda_m\Big(\sqrt{\sum_{k} \gamma_i^k\abs{\braket{i^k}{\psi_m}}^2}\sqrt{\sum_k \gamma_i^k}\Big)\\
&\times\Big(\sqrt{\sum_{k'}\gamma_{i'}^{k'}\abs{\braket{i'^{k'}}{\psi_m}}^2}\sqrt{\sum_{k'} \gamma_{i'}^{k'}}\Big)\\
&=\sum_{i,i'}\sqrt{y_{i}y_{i'}V_iV_{i'}}\sum_m\lambda_m\norm{\vec{a}_i^m}_2\norm{\vec{a}_{i'}^m}_2\\
&\leq \sum_{i,i'}\sqrt{y_{i}y_{i'}V_iV_{i'}}\sqrt{\sum_m\lambda_m\norm{\vec{a}_i^m}_2^2}\sqrt{\sum_{m'}\lambda_{m'}\norm{\vec{a}_{i'}^{m'}}_2^2}\\
&= \sum_{i,i'}\sqrt{p_ip_{i'}y_{i}y_{i'}V_iV_{i'}}= \Big(\sum_{i}\sqrt{p_iy_{i}V_i}\Big)^2.
\end{split}
\]
where $V_i=\sum_k \gamma_i^k$, $p_i=\sum_m\lambda_m\norm{\vec{a}_i^m}_2^2$, $\vec{a}_i^m=(\sqrt{\gamma_i^1}\abs{\braket{i^1}{\psi_m}},\sqrt{\gamma_i^2}\abs{\braket{i^{2}}{\psi_m}},\dots)$, and
$\norm{~}_2$ is the two-norm. The first inequality follows from the definition of $y_i$. We used the Cauchy-Schwarz inequality of type $\abs{\sum_i a_i b_i}\leq \sqrt{\sum_i a_i^2}\sqrt{\sum_{i'} b_{i'}^2}$ in the second and the third inequality.

Combining the two bounds, we obtain Eq.~\eqref{eq:lower_bound_methods}.
\end{proof}

\sect{Proof of collective bounds. }\label{app:bound_collections}
Finally, we prove
\[
\bv^T\bA_i\bv\ \leq\  p_i,
\]
where $\bv_E=\sqrt{p_E}$ and $(\bA_i)_{EE'}=(-1)^{1+\delta_{EE'}}\abs{\bra{E}\Pii_i\ket{E'}}$. Expanding this gives $\sum_{E,E'}\sqrt{p_E}(\bA_{i})_{EE'}\sqrt{p_{E'}}\leq p_i$.
\begin{proof}
\[
\begin{split}
p_i&=\tr[\Pii_i\R]=\sum_{E,E'}\tr[\pro{E}{E}\Pii_i\pro{E'}{E'}\R]\\
&=\sum_{E,E'}\bra{E}\Pii_i\ket{E'}\bra{E'}\R\ket{E}\\
&=\sum_{E}\bra{E}\Pii_i\ket{E}p_E+\sum_{E\neq E'}\bra{E}\Pii_i\ket{E'}\bra{E'}\R\ket{E}\\
&\geq\sum_{E}\bra{E}\Pii_i\ket{E}p_E-\sum_{E\neq E'}\abs{\bra{E}\Pii_i\ket{E'}}\abs{\bra{E'}\R\ket{E}}\\
&\geq\sum_{E}\bra{E}\Pii_i\ket{E}p_E-\sum_{E\neq E'}\abs{\bra{E}\Pii_i\ket{E'}}\sqrt{p_Ep_{E'}}\\
&=2\sum_{E}\bra{E}\Pii_i\ket{E}p_E-\sum_{E,E'}\abs{\bra{E}\Pii_i\ket{E'}}\sqrt{p_Ep_{E'}}\\
&=\sum_{E,E'}\sqrt{p_E}(\bA_{i})_{EE'}\sqrt{p_{E'}},
\end{split}
\]
where $p_E=\bra{E}\R\ket{E}$. For the last inequality, we have used the fact that $\R$ is positive semi-definite, and therefore according to Sylvester's criterion for positive semi-definite matrices~\cite{swamy1973sylvester}, which says that all submatrices must have a non-negative determinant (i.e., all the principal minors are non-negative). For $2$ by $2$ submatrices this means
\[
\bra{E}\R\ket{E}\bra{E'}\R\ket{E'}-\bra{E}\R\ket{E'}\bra{E'}\R\ket{E}\geq 0,
\]
and thus $\abs{\bra{E'}\R\ket{E}}^2\leq p_Ep_{E'}$. The second inequality, $\bv^T\bB_i\bv\ \geq\  p_i$, is proved analogously.
\end{proof}

\sect{Derivation of the analytic bound on the mean energy.} 
Here we derive the recurrence formula for the computation of the lower bound upper values of the bound
\[\label{eq:bounds_on_observable_app}
E_{\min}^{\mathrm{lin}}\leq \mean{E}\leq E_{\max}^{\mathrm{lin}}.
\]
We do this with the lower value, $E_{\min}^{\mathrm{lin}}$, and the formula for the upper value with follow analogously.

The lower bound is given by 
\[\label{eq:newlower_bound_app}
E_{\min}^{\mathrm{lin}}
=\sum_{l=1}^{N} E_l u_l.
\] 
(Eq.~\eqref{eq:newlower_bound}
in the main text), where $u_l$ follow a recurrence relation that we derive next. We simplified the lower indices as $l\equiv E_l$.

The idea behind the recurrence relation is described in Fig.~\ref{Fig:bound} in the main text. Having bounds
\[\label{eq:linear_inequalities_app}
a_{l}^{\max}\equiv a_{E_l}^{\max}\leq p_{E_l} \leq b_{E_l}^{\min}\equiv b_{l}^{\min},
\]
(Eq.~\eqref{eq:linear_inequalities}
 in the main text) we find the lower bound on the mean energy by filling the minimal probability of each energy given by $a_l^{\max}$ and topping it up to the maximum $b_l^{\min}$, from the lowest to the highest energy, until the probabilities sum up to one.

What does this mean mathematically? Let us think of a ``bottle'' of probabilities with the total volume equal to one, $V=1$. We start with all $u_l$ initialized at zero. We pour the minimal required amount given by  $a_E^{\max}\leq p_E$ into each probability $u_l$. After this, we have
\[
\begin{split}
u_1&=a_1^{\max}\\
u_2&=a_2^{\max}\\
&\cdots\\
u_N&=a_N^{\max}\\
\end{split}
\]
and the remaining volume in the bottle is $V=1-\sum_{l=1}^N a_l^{\max}$. We start topping each $u_l$ up to its maximum value, from the lowest to the highest energy eigenvalue, until we run out of the probability in the bottle. For $l=1$, the two cases can occur: either there is enough probability in the bottle to fill $u_l$ up to its maximum allowed value $b_1^{\min}$, or not. Mathematically, this topping-up is expressed as
\[
u_1=a_1^{\max}+\min \left\{b_1^{\min}-a_1^{\max},1-\sum_{l=1}^N a_l^{\max}\right\}.
\]
The $a_1^{\max}$ can be subtracted, which gives
\[
u_1=\min \left\{b_1^{\min},1-\sum_{l=2}^N a_l^{\max}\right\}.
\]
The remaining volume in the bottle is $V=1-u_1-\sum_{l=2}^N a_l^{\max}$.

Next, we top up the second probability, which gives
\[
\begin{split}
u_2&=a_2^{\max}+\min \left\{b_2^{\min}-a_2^{\max},1-u_1-\sum_{l=2}^N a_l^{\max}\right\}\\
&=\min \left\{b_2^{\min},1-u_1-\sum_{l=3}^N a_l^{\max}\right\}.
\end{split}
\]
The remaining volume is $V=1-\sum_{l=1}^2 u_l-\sum_{l=3}^N a_l^{\max}$. We continue up to the maximal index $N$, deriving the full recursive relation, Eq.~\eqref{eq:recursiveprob} 
in the main text.

The recursive relation for $E_{\max}^{\mathrm{lin}}$ is derived analogously.

\section{Powerful improvement from collective bounds}\label{app:considerable_improvement}

Here we show an example in which the collective bounds
\[\label{eq:sum_inequalities_extremized_app}
\bv^T\bA_i\bv\ \leq\  p_i^{\min}\ \leq p_i^{\max}\ \leq\ \bv^T\bB_i\bv,
\]
(Eq.~\eqref{eq:sum_inequalities_extremized}
 in the main text) where 
\[
p_i^{\min}=\min_{t\in [0,T]} p_i(t),\quad p_i^{\max}=\max_{t\in [0,T]} p_i(t),
\]
provide a considerable improvement in the estimation of energy probabilities in comparison with 
using just the linear bound
\[\label{eq:linear_inequalities_app2}
a_E^{\max}\leq p_E \leq b_E^{\min}.
\]
(Eq.~\eqref{eq:linear_inequalities}
in the main text.)

Consider a coarse-grained energy measurement $\C=\{\P_{\tilde E}\}$ given by the coarse-grained energy projectors 
\[\label{eq:coarseenergyp}
\P_{\tilde E}=\sum_{E\in [\tilde E,\tilde E+\Delta E)}\pro{E}{E}.
\]
$\Delta E$ denotes the resolution in measuring energy. Then $\bA_{\tilde E}=\bB_{\tilde E}$ are diagonal, and the inequalities Eq.~\eqref{eq:sum_inequalities_extremized_app} yield
\[\label{eq:coarseprobabilitiesenergy}
p_{\tilde E}=p_{\tilde E}^{\max}=p_{\tilde E}^{\min}= \sum_{E\in [\tilde E,\tilde E+\Delta E)}p_{E}.
\]
This upper bounds the sum of energy probabilities, making the determination of the mean energy much more precise. This is a stark difference with Eq.~\eqref{eq:linear_inequalities_app2}, which yields much less restrictive $p_E\leq p_{\tilde E}$ for each $E\in [\tilde E,\tilde E+\Delta E)$.

To give an example, consider a Hamiltonian $\ham=\sum_{i=1}^8 E_i\pro{E_i}{E_i}$ with the following spectrum
\[
\{E_1,\dots,E_8\}=\{0,\,1,\,2,\,2.5,\,3,\,3.3,\,3.7,\,4\}.
\]
Consider a fixed resolution in measuring energy to be $\Delta E=1$. This results in coarse-grained energy projectors, Eq.~\eqref{eq:coarseenergyp}, representing the coarse energy measurement $\C=\{\P_{\tilde E_j}\}_{j=1}^5$ as
\begin{align}
\P_{\tilde E_1}&=\pro{0}{0},\nonumber\\
\P_{\tilde E_2}&=\pro{1}{1},\nonumber\\
\P_{\tilde E_3}&=\pro{2}{2}+\pro{2.5}{2.5},\\
\P_{\tilde E_4}&=\pro{3}{3}+\pro{3.3}{3.3}+\pro{3.7}{3.7},\nonumber\\
\P_{\tilde E_5}&=\pro{4}{4}.\nonumber
\end{align}
This indicates that the measurement device cannot distinguish between energy states $\ket{2}$ and $\ket{2.5}$, for example, because they are too close in energy.

Consider an initial state
\[
\ket{\psi}=(\ket{2}+\ket{3})/\sqrt{2}.
\]
Knowing this state, we can compute
\[
\begin{split}
p_{0}&=0,\ p_{1}=0,\ p_{2}=1/2,\ p_{2.5}=0,\\
p_{3}&=1/2,\ p_{3.3}=0,\ p_{3.7}=0,\ p_{4}=0.
\end{split}
\]
However, the experimenter performing a coarse-grained measurement $\C$ on many copies of this initial state does not know that. Instead, using Eqs.~\eqref{eq:linear_inequalities_app2}, they derive
\[
\begin{split}
p_{0}&=0,\ p_{1}=0,\ p_{2}\leq 1/2,\ p_{2.5}\leq 1/2,\\
p_{3}&\leq 1/2,\ p_{3.3}\leq 1/2,\ p_{3.7}\leq 1/2,\ p_{4}=0.
\end{split}
\]
The right-hand sides were obtained from the outcomes of the coarse-grained measurement. This is because half of the time, they obtain measurement outcome $\tilde E_3$, and the other half they get $\tilde E_4$. If they were to estimate the mean energy of the state only from these equations, they would obtain
\[
2\times\frac{1}{2}+2.5\times\frac{1}{2}=2.25\leq \mean{E}\leq3.5= 3.3\times\frac{1}{2}+3.7\times\frac{1}{2}.
\]
However, from Eqs.~\eqref{eq:coarseprobabilitiesenergy}, which follow from the collective bounds, they obtain an additional set of equations,
\[
\frac{1}{2}=p_{2}+p_{2.5},\quad \frac{1}{2}=p_{3}+p_{3.3}+p_{3.7}.
\]
The left-hand sides were obtained from the experimental outcomes. Using this additional set of equations, the experimenter is able to derive a noticeably tighter bound on the mean energy,
\[
2\times\frac{1}{2}+3\times\frac{1}{2}=2.5\leq \mean{E}\leq3.1= 2.5\times\frac{1}{2}+3.7\times\frac{1}{2}.
\]
This improvement will be dramatic in systems with many energy eigenstates, leading to much coarser projectors.

\section{Quality factors}\label{app:quality}

We can employ two quality factors to assess the performance of method four bounding the mean energy: the first,
\[\label{eq:Q1}
Q_1=\left(1-\frac{E_{\max}-E_{\min}}{E_N-E_1}\right)\times 100\%,
\]
which measures the range of excluded energy, and the second,
\[\label{eq:Q2}
Q_2=\left(1-\frac{N_{[E_{\min},E_{\max}]}}{N}\right)\times 100\%,
\]
which measures the percentage of ``excluded'' energy eigenstates. $N_{[E_{\min},E_{\max}]}$ denotes the number of energy eigenstates with energy between $E_{\min}$ and $E_{\max}$, and $N$ is the dimension of the Hilbert space.

\begin{figure}[t!]
\begin{center}
\includegraphics[width=3.5cm]{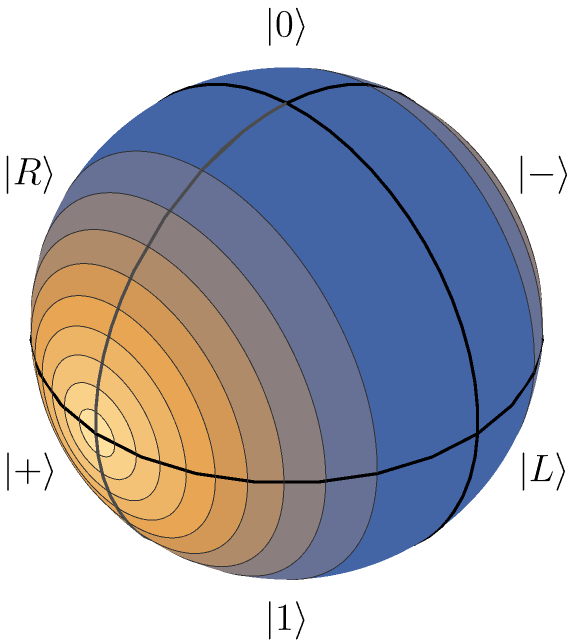}
\includegraphics[width=4.4cm]{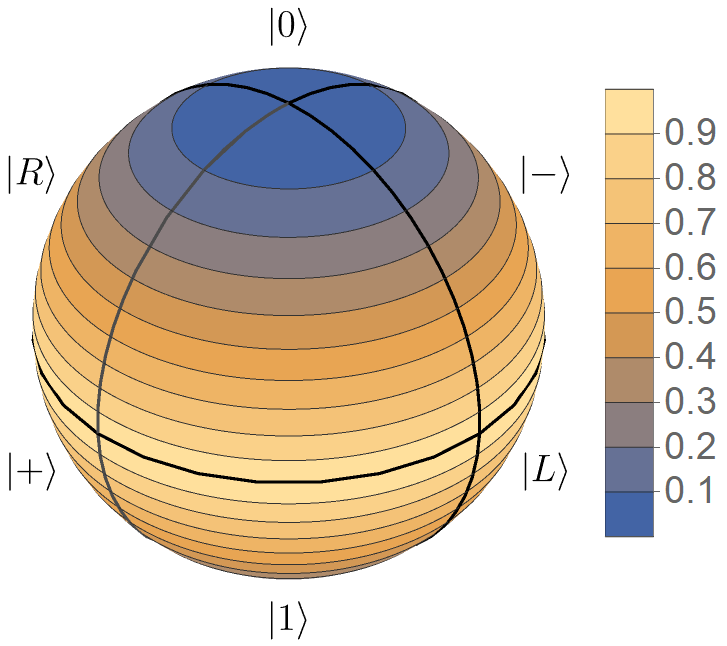}
\\
\caption
{Illustration of the performance when estimating energy by local Pauli-$x$ measurements in a simple qubit example. We show the quality $Q_1$ factor---percentage of excluded energies---when estimating energy given by the Hamiltonian $\ham=\hat\sigma_z$ by measuring in the $\hat\sigma_x$ basis, for different initial states on the Bloch sphere. Left panel: Quality factor of the initial state. Value $Q_1=1$ (light orange; which happens for state $\ket{\psi}=\ket{+}$) corresponds to perfect identification of the energy, while value $Q_1=0$ (dark blue; which happens for state $\ket{\psi}=\ket{0}$) corresponds to failure in identifying the energy. Right panel: the quality factor over initial states if we allow time evolution of the state with the Hamiltonian and measure at different times $t\in [0,\pi/2)$ to build up statistics of outcomes at each time of this interval. The Hamiltonian revolves the state around the z-axis, so during the time evolution, the state picks up the best quality factor at that latitude from the figure on the left.}
\label{Fig:bloch}
\end{center}
\end{figure}

\section{Simple example}\label{app:simple}

Consider a Hamiltonian given by the Pauli-z Matrix,
\[
\ham=\hat{\sigma}_z=\begin{pmatrix}
1 & 0 \\
0 & -1
\end{pmatrix},
\]
which has energy eigenvalues $E_0=-1$ and $E_1=1$, corresponding to eigenstates $\ket{0}$ and $\ket{1}$, respectively. The task is to estimate the energy of a general pure qubit state,
\[
\ket{\psi}=\cos \frac{\theta}{2} \ket{0}+e^{i \phi}\sin \frac{\theta}{2} \ket{1},
\]
where $0
\leq \theta \leq \pi$ and $0
\leq \phi \leq 2\pi$.

We consider two different two-outcome measurements to estimate energy. We will use the combined bound for the energy probabilities,
\[
\abs{p_E-\sum_i p_i\abs{\braket{E}{i}}^2}\leq b_E-\sum_i p_i\abs{\braket{E}{i}}^2,
\]
where $b_E=\big(\sum_i\sqrt{p_i}\abs{\braket{E}{i}}\big)^2$, which is easily derived from Eq.~\eqref{eq:time_dependent_bound_ab}
in the main text.

First, consider measuring in the z-basis, i.e., measuring in the eigenbasis of the operator $\hat{M}=\hat\sigma_z$. This defines the measurement $\C=\{\pro{0}{0},\pro{1}{1}\}$. Because the measurement basis is the same as the eigenbasis of the Hamiltonian, we are measuring the energy directly, so we expect the exact result. From the bound above, we have
\[
\begin{split}
    \abs{p_{E_0}-p_0}&\leq 0,\\
    \abs{p_{E_1}-p_1}&\leq 0,
\end{split}
\]
independent of the initial state $\ket{\psi}$. Clearly, $p_{E_0}=p_0$, $p_{E_1}=p_1$, and $E_{\min}^{\mathrm{lin}}=\mean{E}=E_{\max}^{\mathrm{lin}}$, as expected.

Second, consider measuring in the x-basis, i.e., measuring in the eigenbasis of the operator $\hat{M}=\hat \sigma_x$. This defines $\C=\{\pro{+}{+},\pro{-}{-}\}$. We have
\[
\begin{split}
    \abs{p_{E_0}-\tfrac{1}{2}}&\leq \tfrac{1}{2}\big((\sqrt{p_+}+\sqrt{p_-})^2-1\big),\\
    \abs{p_{E_1}-\tfrac{1}{2}}&\leq \tfrac{1}{2}\big((\sqrt{p_+}+\sqrt{p_-})^2-1\big).
\end{split}
\]
This means that if the state is aligned with the $x$ axis, for example, $p_+=1$, then $p_{E_0}=p_{E_1}=\tfrac{1}{2}$ and we can determine the energy exactly as $\mean{E}=0$. On the contrary, if the state is aligned with the z-axis, implying $p_+=p_-=\frac{1}{2}$, then $\abs{p_{E_0}-\tfrac{1}{2}}\leq\tfrac{1}{2}$ and $\abs{p_{E_1}-\tfrac{1}{2}}\leq\tfrac{1}{2}$ which we can rewrite as $0\leq p_{E_0} \leq 1$ and $0\leq p_{E_1} \leq 1$. Thus we obtain a trivial bound, 
\[
-1=E_{\max}^{\mathrm{lin}}\leq \mean{E} \leq E_{\max}^{\mathrm{lin}}=1.
\]
Generally, we have
\[
p_+=\tfrac{1}{2}(1+\cos \phi \sin \theta),\quad p_-=\tfrac{1}{2}(1+\cos \phi \sin \theta).
\]
which gives
\[
\begin{split}
    \abs{p_{E_0}-\tfrac{1}{2}}&\leq \tfrac{1}{2}\sqrt{1-\cos^2 \phi \sin^2 \theta},\\
    \abs{p_{E_1}-\tfrac{1}{2}}&\leq \tfrac{1}{2}\sqrt{1-\cos^2 \phi \sin^2 \theta}.
\end{split}
\]
This yields
\[
-\!\sqrt{1\!-\!\cos^2 \!\phi \sin^2 \!\theta}\!=\!E_{\max}^{\mathrm{lin}}\!\leq \mean{E} \leq\! E_{\max}^{\mathrm{lin}}\!=\!\sqrt{1\!-\!\cos^2 \!\phi \sin^2 \!\theta}.
\]
We visualize the corresponding quality factor (the percentage of excluded energies, see Eq.~\eqref{eq:Q1})
\[
Q_1=1-\sqrt{1\!-\!\cos^2 \!\phi \sin^2 \!\theta}
\]
for this general case on the Bloch sphere in Fig.~\ref{Fig:bloch}.

Next, we consider time evolution. We have
\[
\ket{\psi_t}=e^{-i \ham t}\ket{\psi}=\cos \frac{\theta}{2} \ket{0}+e^{i (\phi-2t)}\sin \frac{\theta}{2} \ket{1}.
\]
Measuring $\hat \sigma_x$ at time $t$ bounds the energy probabilities as
\[
\begin{split}
    \abs{p_{E_0}-\tfrac{1}{2}}&\leq \tfrac{1}{2}\sqrt{1-\cos^2 (\phi-2t) \sin^2 \theta},\\
    \abs{p_{E_1}-\tfrac{1}{2}}&\leq \tfrac{1}{2}\sqrt{1-\cos^2 (\phi-2t) \sin^2 \theta}.
\end{split}
\]
If we measure at different times during $t\in [0,\pi/2)$, we manage to tighten these bounds and obtain
\[
\begin{split}
    \abs{p_{E_0}-\tfrac{1}{2}}&\leq \tfrac{1}{2}\abs{\cos \theta},\\
    \abs{p_{E_1}-\tfrac{1}{2}}&\leq \tfrac{1}{2}\abs{\cos \theta},
\end{split}
\]
which leads to bounds on energy
\[
-\abs{\cos \theta}=E_{\max}^{\mathrm{lin}}\leq \mean{E} \leq E_{\max}^{\mathrm{lin}}=\abs{\cos \theta}.
\]
The corresponding quality factor $Q_1=1-\abs{\cos \theta}$ is again plotted in Fig.~\ref{Fig:bloch}.

\section{Estimation of the mean energy in experimentally relevant models}\label{app:experimental}

Here we show the simulations for energy estimation using measurements in a local number basis and then using optimized $k$-local measurements. This is a continuation of numerical simulations shown in the main text, which contained a part of the results obtained for the Heisenberg model.

We simulate a number of models, including the Heisenberg model~\cite{porras2004effective,pal2010many-body, luitz2015many}, which is a paradigmatic model to study for many-body localization, and then several other experimentally relevant models. These are the Ising model~\cite{smith2016many,jurcevic2017direct,zhang2017observation,bingham2021experimental}, known for its frequent use in quantum simulators and quantum annealers, the XY model~\cite{lanyon2017efficient,friis2018observation,brydges2019probing,maier2019environment}, which is a type of non-integrable long-range model, and the PXP model~\cite{bernien2017probing,Turner2018, su2022observation}, an archetypal model for many-body quantum scars.

The Hamiltonians and the corresponding parameters are given in Table~\ref{tab:hamiltonians}. The simulations of energy estimation using local particle number measurements and optimized $k$-local measurements are shown in Tables~\ref{tab:plots_app} and~\ref{tab:plots_app2}.

The bulk of the explanation necessary to understand these numerical experiments are also shown in the main text. Below, we give details on the types of Hilbert space considered in our simulations, and we discuss methods that we designed for the optimization over the $k$-local measurements to estimate energy.

\sect{Hilbert space considered.}
In our simulations, we choose to work in a different type of Hilbert space for each model, depending on the conservation laws, and to match the experimental setups, see Table~\ref{tab:hamiltonians}: In the Heisenberg and the XY models, the full Hilbert space splits into subspaces, each of them characterized by a definite value of the total spin along the $z$ axes, i.e., $\hat S^z = \sum_{i} \hat \sigma_i^z$. We work in the largest subspace, characterized by the value $\hat{S}^z = 0$. 
This conservation is also why the actual value of $B$ in the XY model is irrelevant. 
For the Ising model, the total spin along the $z$ axes is not conserved, only the \textit{parity} of the total spin is conserved.
In this case, we work on the parity even subspace, which contains the N\'eel state. For the PXP model, the situation is more intricate: the presence of the projector operators $(\hat{I} - \hat{\sigma}_i^z)$ in the Hamiltonian introduces a non-trivial local constraint \cite{Turner2018}. 
Consequently, the full Hilbert space shatters in many different subspaces, dynamically disconnected and having various dimensions \cite{khemani2020localization}.
Inside each subspace, the dynamics is generically chaotic with the presence of \textit{many-body scars} \cite{Turner2018, serbyn2021quantum, chandran2022quantum}.
However, our goal in using this model was to study the effect of the Hilbert space shattering on the quality of the energy estimation.
Therefore, we work in the full Hilbert space.

\begin{figure}[t!]
\begin{center}
\includegraphics[width=0.34\textwidth]{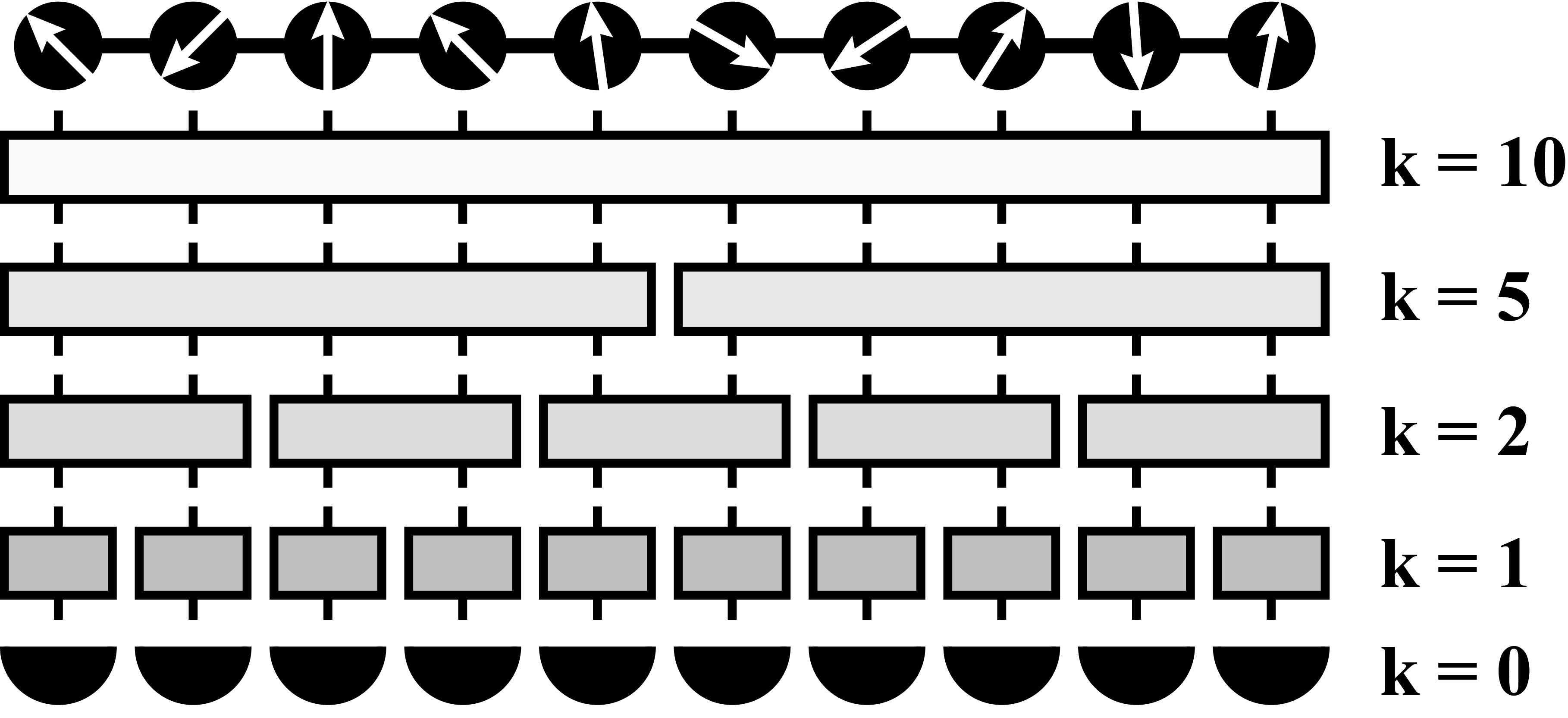}
\\
\caption
{Sketch of allowed operations---$k$-local measurements. $k=0$ corresponds to measuring in the local number basis. $k=1$ allows for using single-site unitary operators leading to a general single-site measurement, $k=2$ for two-site local operators, etc.}
\label{Fig:allowed_op}
\end{center}
\end{figure}

\sect{Methods for optimizing local measurements.} 
We introduce three methods of analytical optimization for the $k$-local measurements. The sketch of $k$-local measurements is shown in Fig.~\ref{Fig:allowed_op}, which corresponds to Fig~\ref{tab:plots_large} (a)
in the main text. $k$-local measurement consists of applying a $k$-local unitary operator and then measuring in the computational basis.

\begin{table*}[ht!]
\begin{tabular}[t]{|c|c|c|c|c|c|c|}
\hline
name & characteristics  & conservation & \multicolumn{2}{c|}{Hamiltonian} & \multicolumn{2}{c|}{parameters}  \\
\hline
\begin{tabular}{c}
Heisenberg \\
(integrable,\\
delocalized, \\
(localized)
\end{tabular} & \begin{tabular}{c}
many-body \\
localizing
\end{tabular} &  \begin{tabular}{c}
total spin\\
conserving
\end{tabular}   & \multicolumn{2}{c|}{$\displaystyle
\ham = \sum_{i} \big(\hat \sigma_i^x\hat \sigma_{i+1}^x+\hat \sigma_i^y\hat \sigma_{i+1}^y+\hat \sigma_i^z\hat \sigma_{i+1}^z\big)+\sum_i h_i\hat \sigma_i^z $} & \multicolumn{2}{c|}{\begin{tabular}{c}
$h_i \in [-W,W]$ drawn randomly\\
($W=0$, $W=0.5$, $W=10$)
\end{tabular}}   \\
\hline
\begin{tabular}{c}
Ising \\
(delocalized,\\
localized) \\
\end{tabular} & \begin{tabular}{c}
gapped and\\
many-body \\
localizing
\end{tabular} &  \begin{tabular}{c}
parity\\
conserving
\end{tabular}   &  \multicolumn{2}{c|}{$\displaystyle
\ham \!=\! \sum_{i<j} J_{ij}\hat \sigma_i^x\hat \sigma_j^x+\frac{1}{2}\sum_i(B+h_i)\hat \sigma_j^z$} & \multicolumn{2}{c|}{\begin{tabular}{c}
$J_0=1$, $\alpha=1.13$, $B=4$, \\
$h_i \in [-W,W]$ drawn randomly \\
($W=0$, $W=8$)
\end{tabular}}   \\
\hline
XY &  \begin{tabular}{c}
gapped and\\
long-range
\end{tabular} &  \begin{tabular}{c}
total spin\\
conserving
\end{tabular}   & \multicolumn{2}{c|}{$\displaystyle
\ham \!=\! \sum_{i<j} J_{ij}\big(\hat \sigma_i^+\hat \sigma_j^-+\hat \sigma_i^-\hat \sigma_j^+\big)+B\sum_i\hat \sigma_i^z$} & \multicolumn{2}{c|}{$J_0 =1 $, $\alpha=1.24$, $B=0$}  \\
\hline
PXP & \begin{tabular}{c}
quantum\\
scars
\end{tabular} &  \begin{tabular}{c}
Hilbert space\\
shattering
\end{tabular}   & \multicolumn{2}{c|}{$\displaystyle
\ham \!=\! \tfrac{\Omega}{4}\sum_{i} \big(\I-\hat\sigma_{i}^z\big) \hat \sigma_{i+1}^x \big(\I-\hat\sigma_{i+2}^z\big)$} & \multicolumn{2}{c|}{$\Omega =1$}  \\
\hline
 & \multicolumn{6}{c|}{Sizes and Hilbert spaces considered in the numerical experiments}  \\
\cline{2-7}
 & \multicolumn{3}{c|}{Small systems} & \multicolumn{3}{c|}{Large systems}   \\
\hline
Heisenberg & \multicolumn{2}{l}{6 sites, 3 particles} & $D=20$ & \multicolumn{2}{l}{10 sites, 5 particles} & \multicolumn{1}{l|}{$D=252$}   \\
Ising & \multicolumn{2}{l}{6 sites, even parity subspace$\quad\quad$} & $D=32$ & \multicolumn{2}{l}{10 sites, even parity subspace$\quad\quad$} & \multicolumn{1}{l|}{$D=512$}   \\
XY & \multicolumn{2}{l}{6 sites, 3 particles} & $D=20$ & \multicolumn{2}{l}{10 sites, 5 particles} & \multicolumn{1}{l|}{$D=252$}   \\
PXP & \multicolumn{2}{l}{5 sites, full Hilbert space} & $D=32$ & \multicolumn{2}{l}{10 sites, full Hilbert space} & \multicolumn{1}{l|}{$D=1024$}   \\
\hline
\end{tabular}
\caption{\label{tab:hamiltonians} Table of models used in our simulations and the Hilbert spaces considered with dimension $D$. $\hat \sigma_i^x$ denotes the Pauli-x matrix at site $i$, and similar with Pauli-y and Pauli-z matrices. $\hat \sigma _i^+$ and $\hat \sigma _i^-$ denote the spin creation and annihilation operators, respectively. The Ising and XY models have a non-local interaction of form $J_{ij}=J_0/\abs{i-j}^\alpha$. Parameters were taken to match those employed in experiments.}
\end{table*}

\begin{table*}[tp!]
\begin{tabular}[t]{|c|c|c|c|}
\hline
\multicolumn{4}{|c|}{initial states: \quad  \color{black}{$\filledstar$} G \quad  {\color{blue}{$\filleddiamond$}} C 
 \quad {\color{red}{$\bullet$}} H}\\
\hline
\multirow{2}{*}{Ham.}  & \multirow{2}{*}{small system} & \multicolumn{2}{c|}{large system} \\
\cline{3-4}
 &    & ground state-optimized &  observable-optimized    \\
\hline
\begin{turn}{90}
\begin{tabular}{cc}
   Heisenberg \\
     (integrable) 
\end{tabular}
\end{turn} & \includegraphics[width=0.3\textwidth]{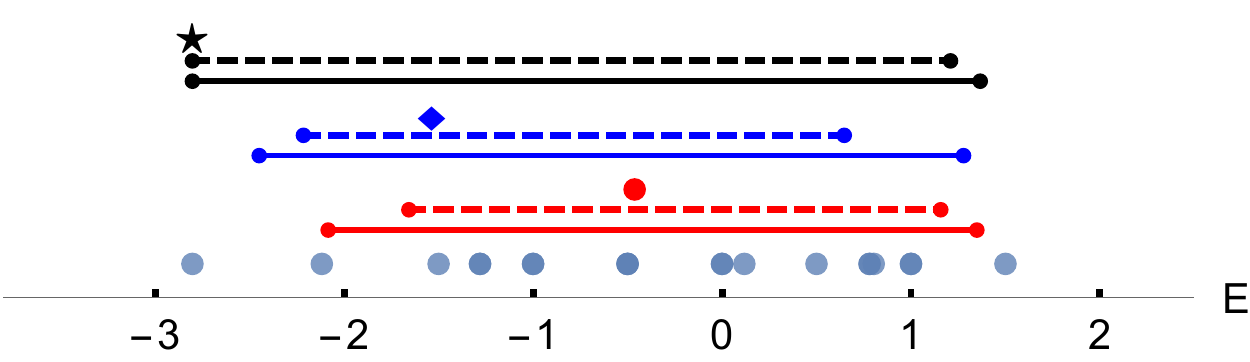} &  \includegraphics[width=0.3\textwidth]{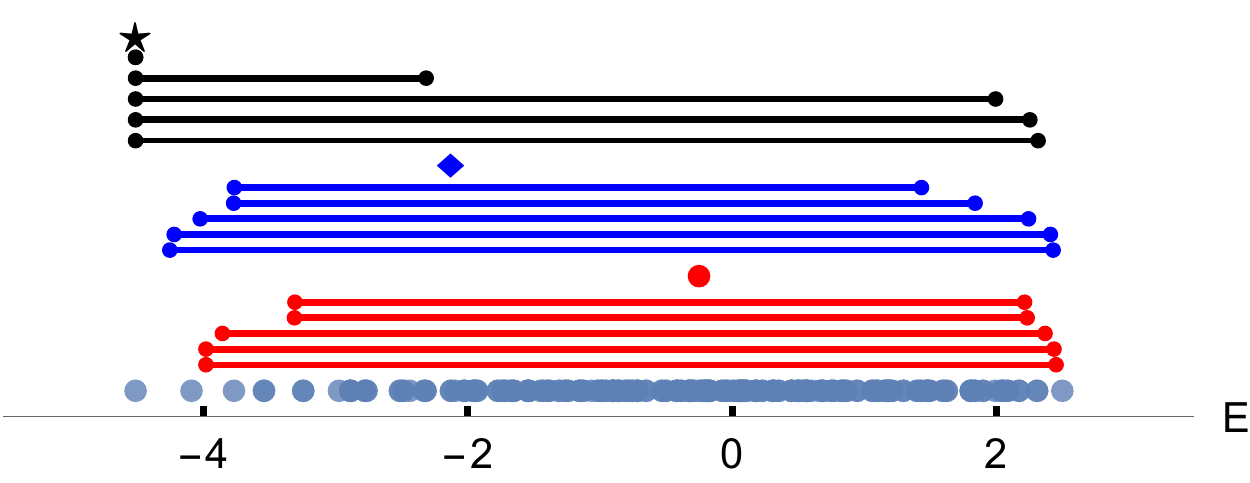}  & \includegraphics[width=0.3\textwidth]{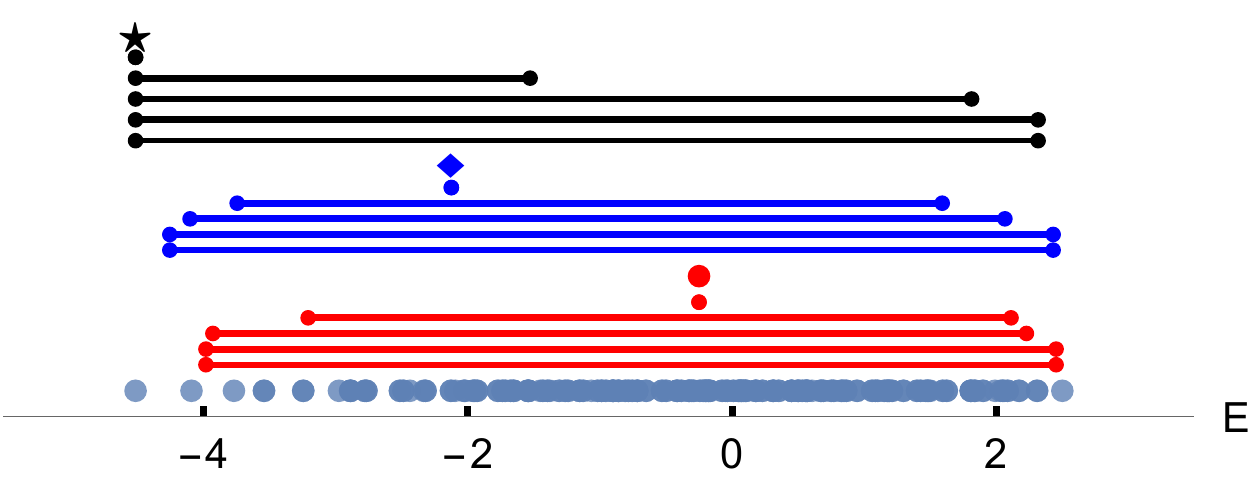}  \\ 
\hline
\begin{turn}{90}
\begin{tabular}{cc}
   Heisenberg \\
     (delocalized) 
\end{tabular}
\end{turn} & \includegraphics[width=0.3\textwidth]{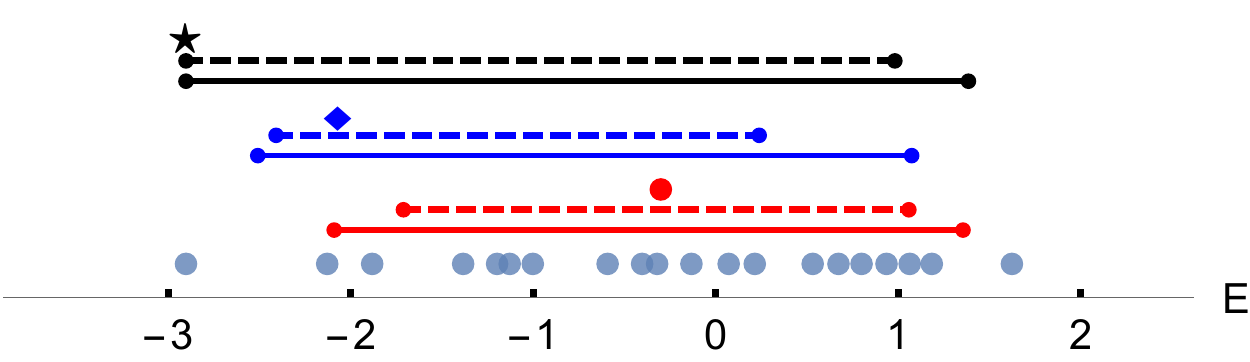} &  \includegraphics[width=0.3\textwidth]{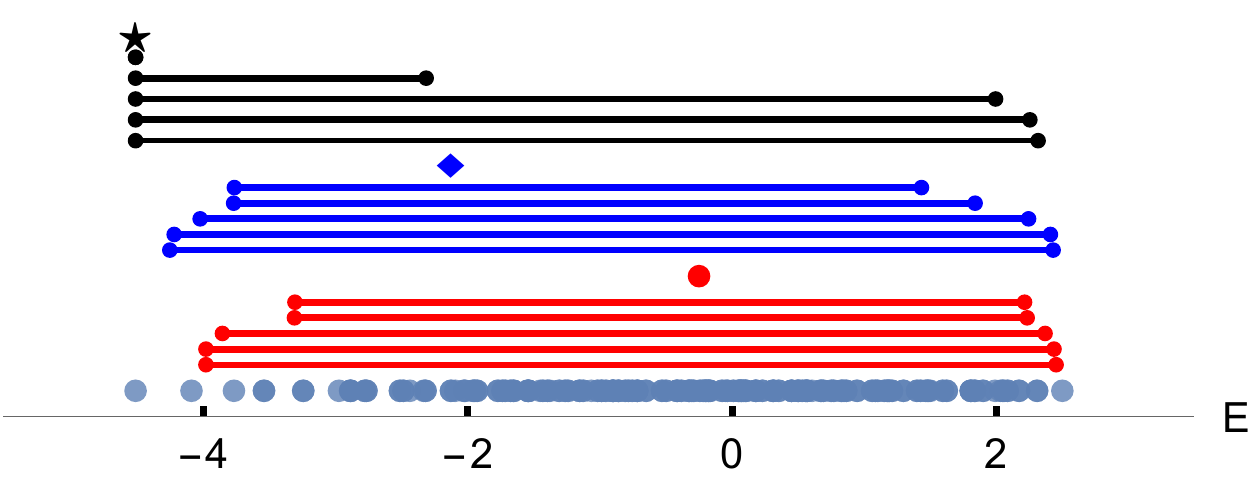}  & \includegraphics[width=0.3\textwidth]{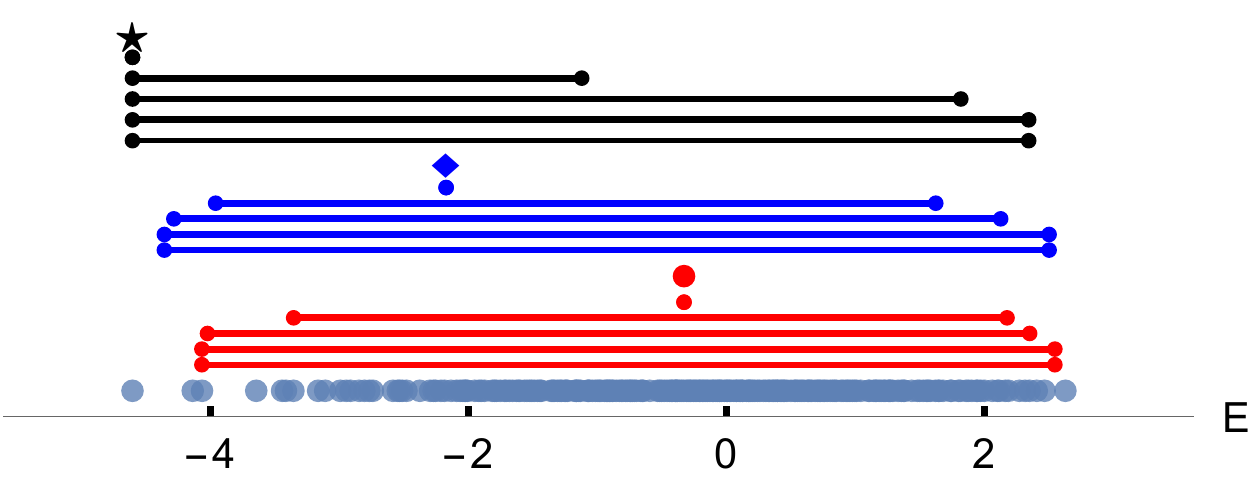}  \\ 
\hline
\begin{turn}{90}
\begin{tabular}{cc}
   Heisenberg \\
     (localized) 
\end{tabular}
\end{turn} & \includegraphics[width=0.3\textwidth]{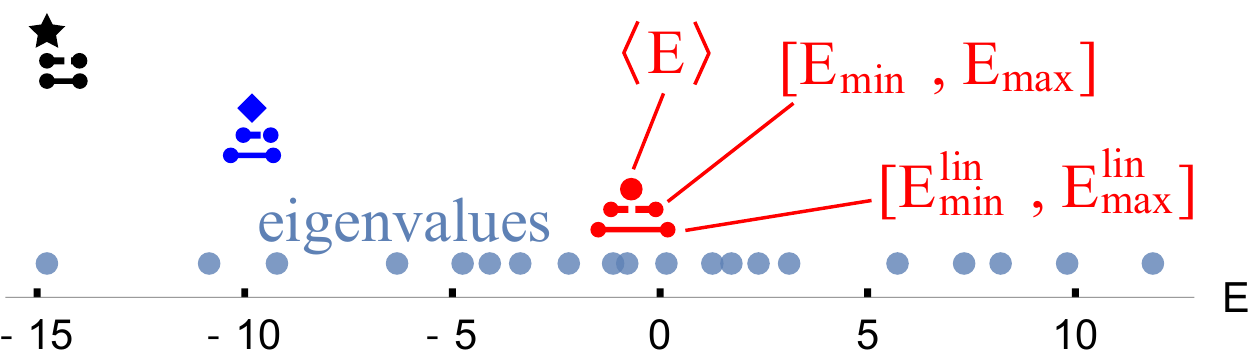} &  \includegraphics[width=0.3\textwidth]{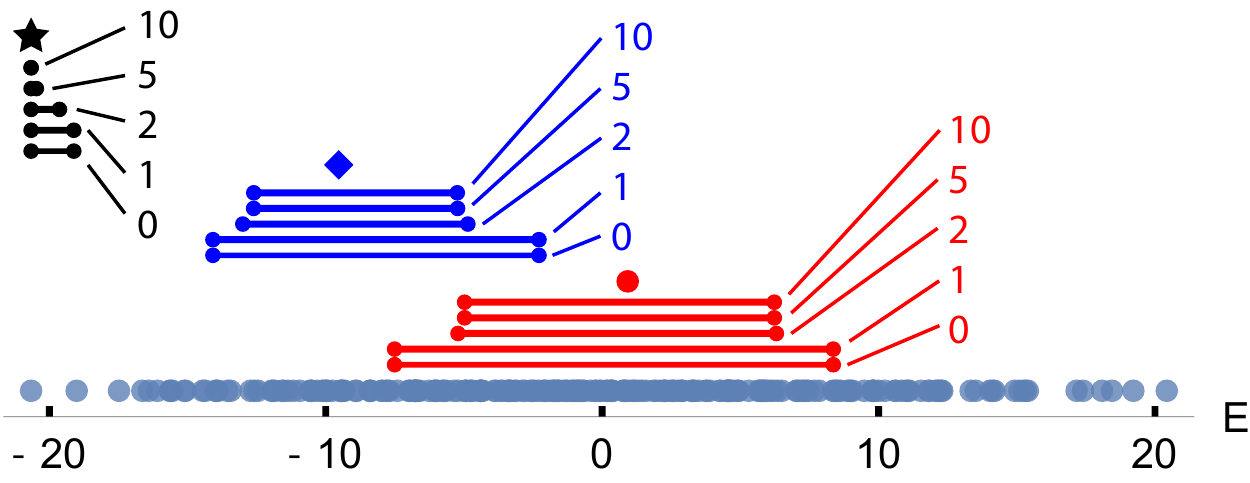}  & \includegraphics[width=0.3\textwidth]{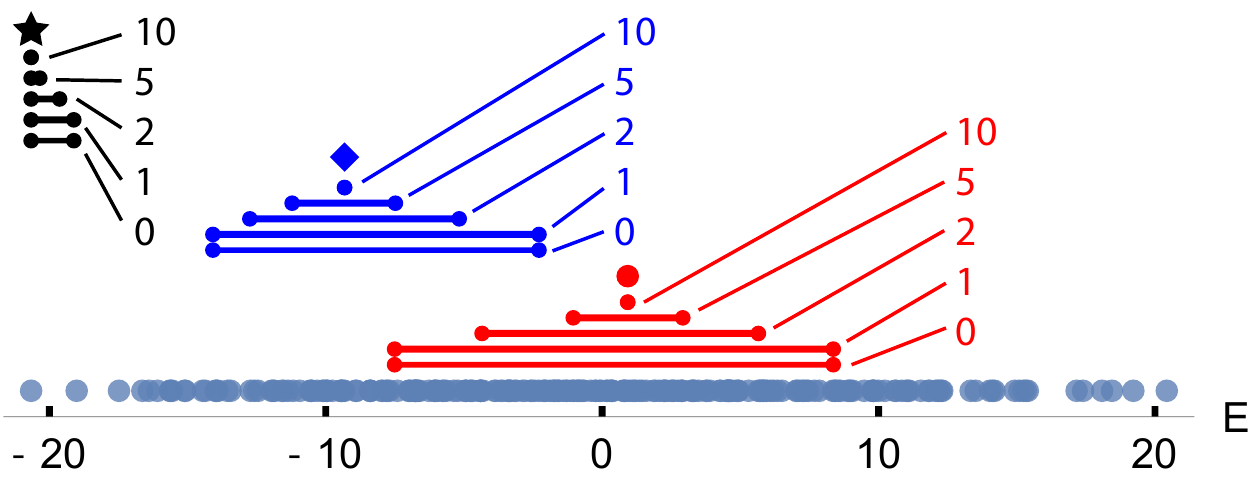}  \\ 
\hline
\begin{turn}{90}
\begin{tabular}{cc}
   Ising \\
     (delocalized) 
\end{tabular}
\end{turn} & \includegraphics[width=0.3\textwidth]{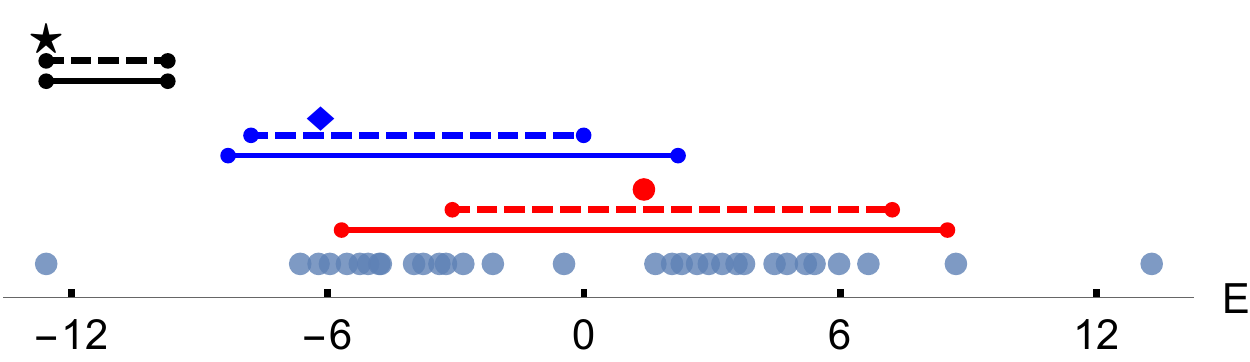} &  \includegraphics[width=0.3\textwidth]{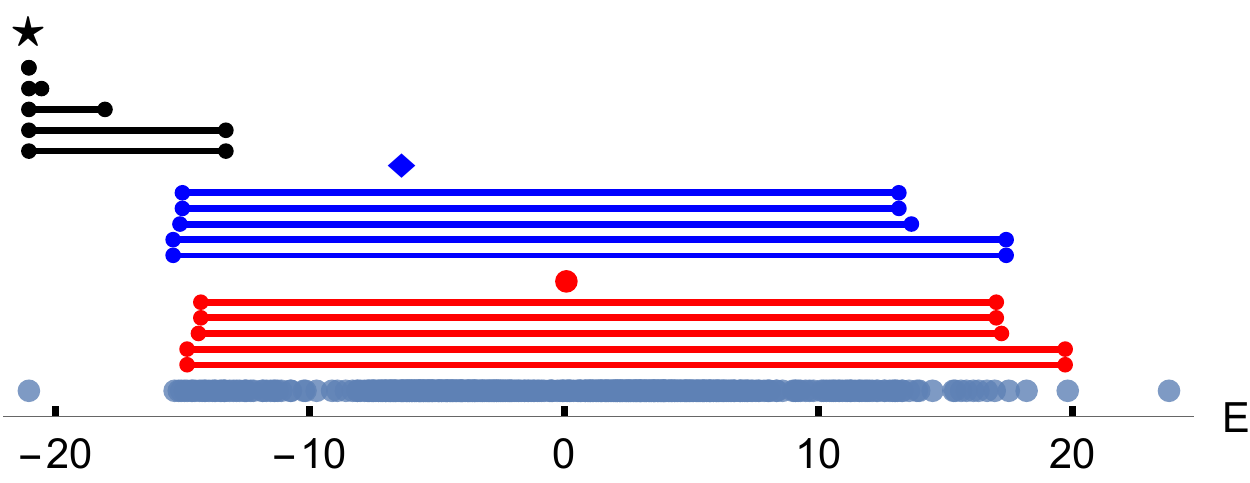}  & \includegraphics[width=0.3\textwidth]{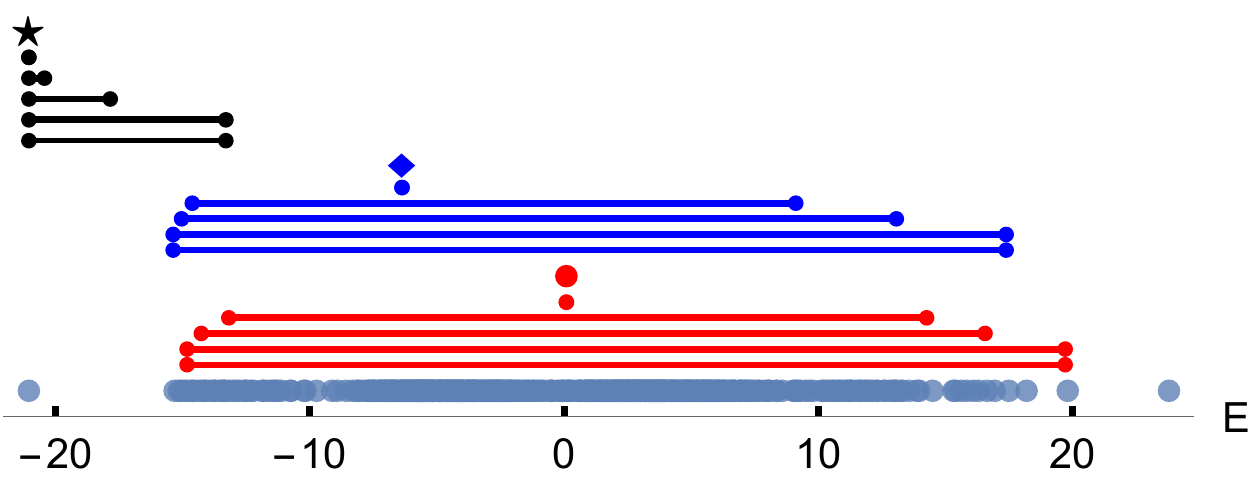}  \\ \hline
\begin{turn}{90}
\begin{tabular}{cc}
   Ising \\
     (localized) 
\end{tabular}
\end{turn} & \includegraphics[width=0.3\textwidth]{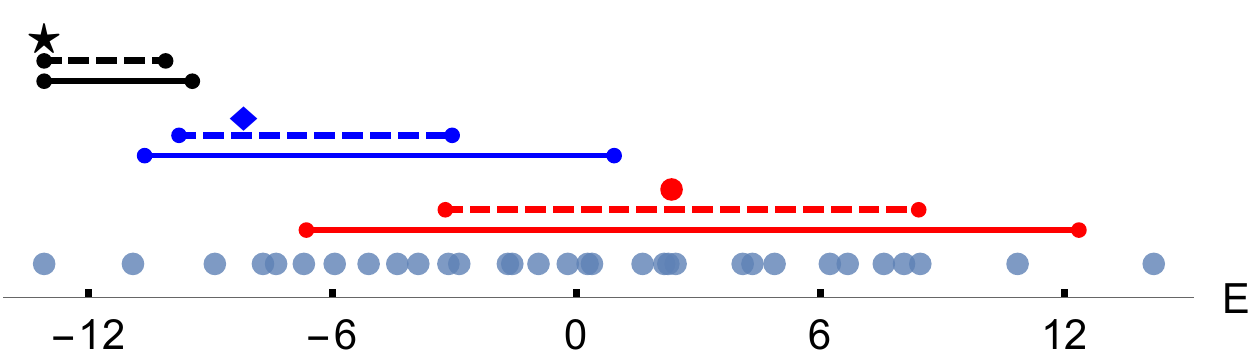} &  \includegraphics[width=0.3\textwidth]{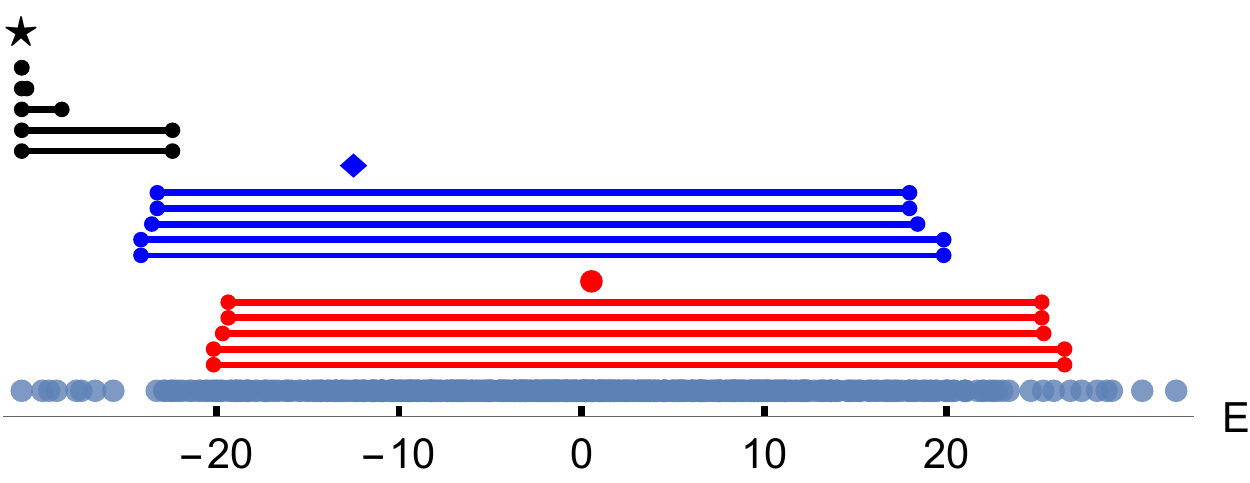}  & \includegraphics[width=0.3\textwidth]{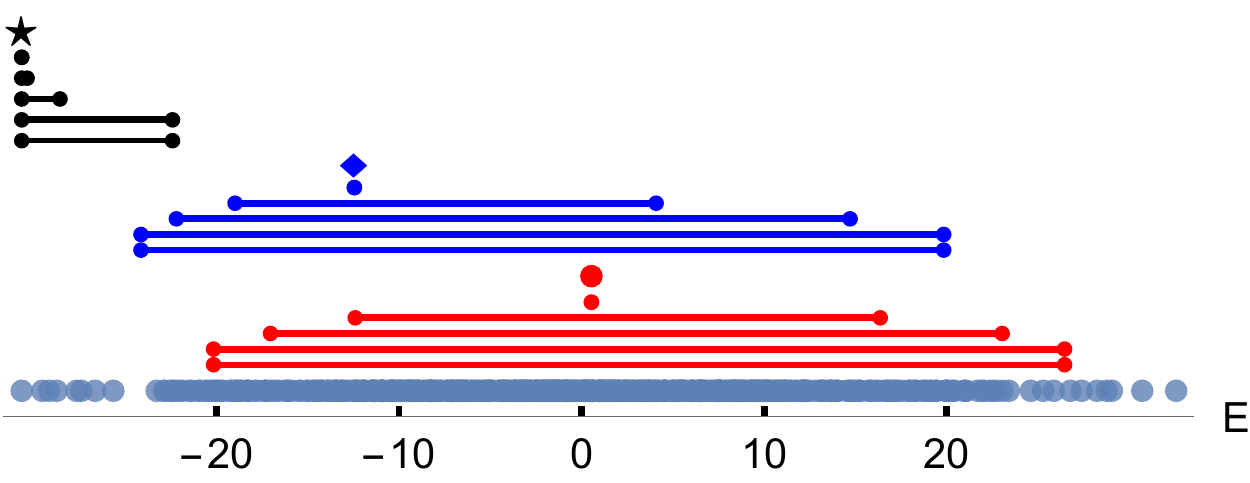}  \\ 
\hline
\end{tabular}
\caption{\label{tab:plots_app} Estimating energy with the local number and $k$-local optimized measurements (see Fig.~\ref{Fig:allowed_op}), for various Hamiltonians and sizes of Hilbert space delineated in Table~\ref{tab:hamiltonians}. The initial state is either a ground state (G), a pure thermal state (C - cold), Eq.~\eqref{eq:initialpurethermal}
in the main text, or a state drawn randomly from the Hilbert space with the Haar measure (H - hot). Small systems (left panel): the graphs show the true mean energy $\mean{E}$ (single symbol), intervals $[E_{\min}^{\mathrm{lin}}, E_{\max}^{\mathrm{lin}}]$ (full-line) and $[E_{\min},E_{\max}]$ (dashed-line), for each state ordered from top to bottom, and the list of energy eigenvalues at the very bottom. Large systems: the true mean energy $\mean{E}$ (single symbol), intervals $[E_{\min}^{{\mathrm{lin}}(k)}, E_{\max}^{{\mathrm{lin}}(k)}]$ (full-lines), denoting analytic bounds computed for $k$-local measurements, $k=0,1,2,5,10$ (see Fig.~\ref{Fig:bloch}), using ground state-optimized method (middle panel), and observable-optimized type 1 method (right panel). See the observable-optimized type 2 method for the PXP model in Fig.~\ref{Fig:PXPtype2}. \textbf{Observations:} 1) There is little difference between the integrable and delocalized phases of the Heisenberg model in estimating energy. Bethe integrability does not seem to play a role. 2) Estimation of energy in the localized phase of the Heisenberg model works well for both small and large systems. It managed to exclude $Q_1=97.5\%$ of the range of energies when estimating the ground state energy using two-qubit ($k=2$) measurements. This is due to a large overlap between energy eigenstates and the local number basis.  3) Estimation of the ground state energy in the Ising model works well both in the localized ($Q_1=96.7\%$ for $k=2$) and delocalized ($Q_1=92.9\%$ for $k=2$) phases. This is due to the low entanglement in the ground state. Please see  the continuation of this table in Table~\ref{tab:plots_app2}.}
\end{table*}

\begin{table*}[tp!]
\begin{tabular}[t]{|c|c|c|c|}
\hline
\begin{turn}{90}
\begin{tabular}{cc}
   XY
\end{tabular}
\end{turn} & \includegraphics[width=0.3\textwidth]{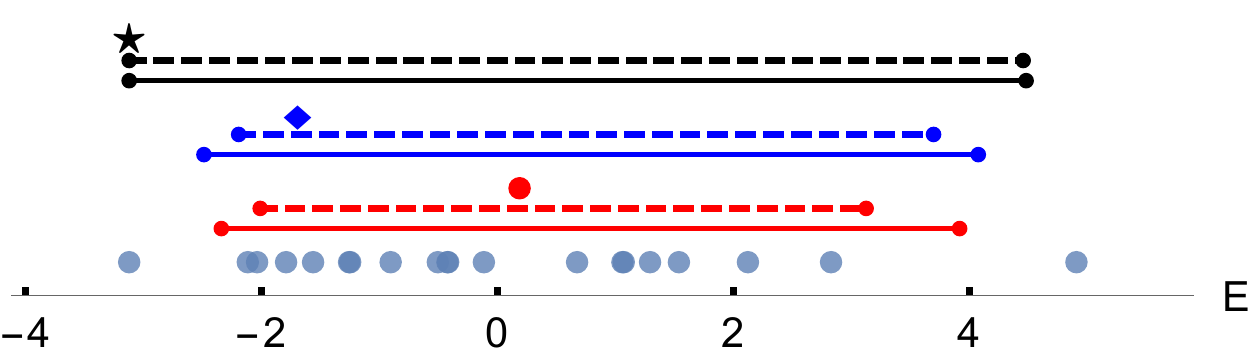} &  \includegraphics[width=0.3\textwidth]{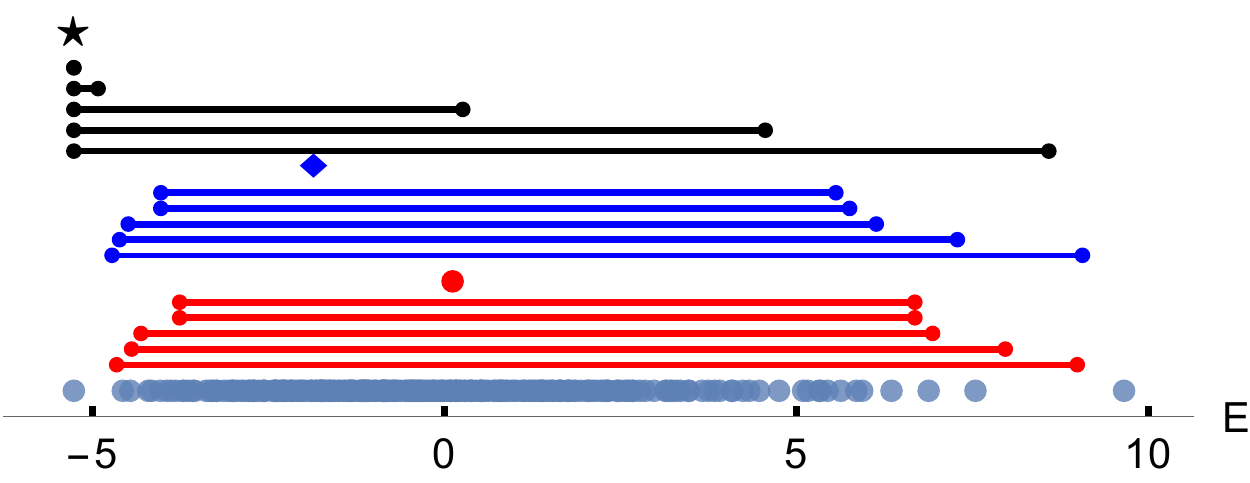}  & \includegraphics[width=0.3\textwidth]{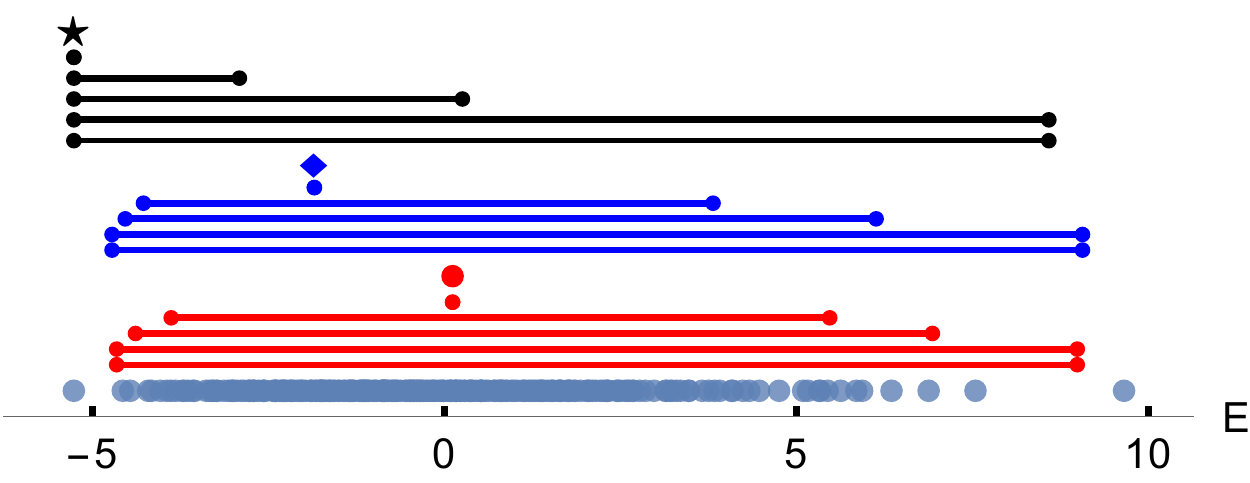}  \\
\hline
\begin{turn}{90}
\begin{tabular}{cc}
PXP
\end{tabular}
\end{turn} & \includegraphics[width=0.3\textwidth]{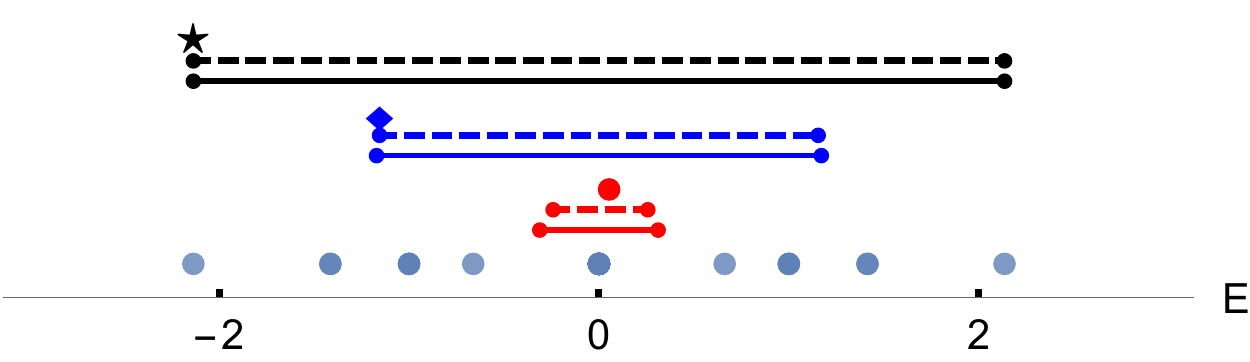} &  \includegraphics[width=0.3\textwidth]{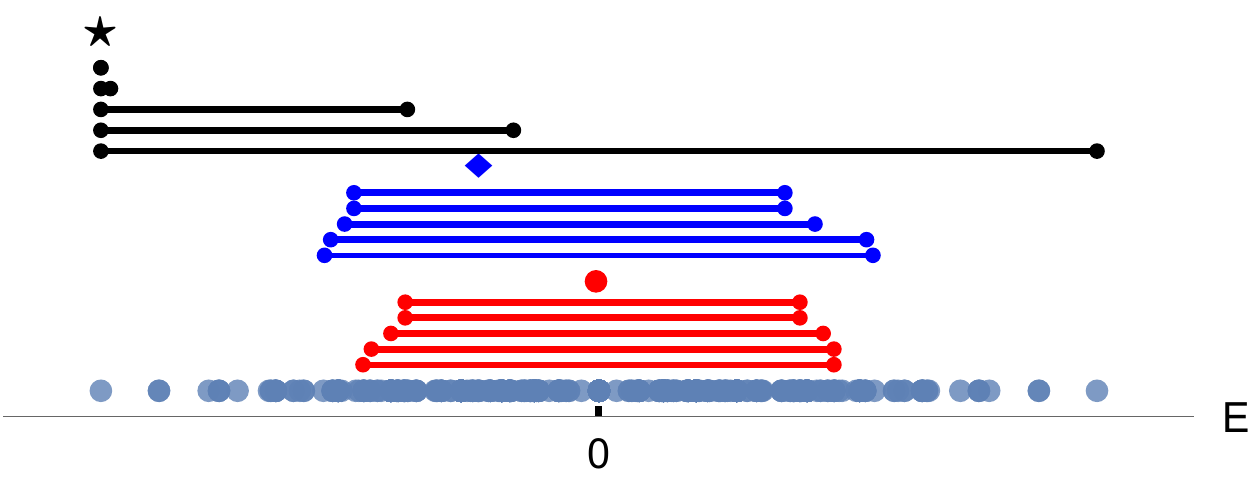}  & \includegraphics[width=0.3\textwidth]{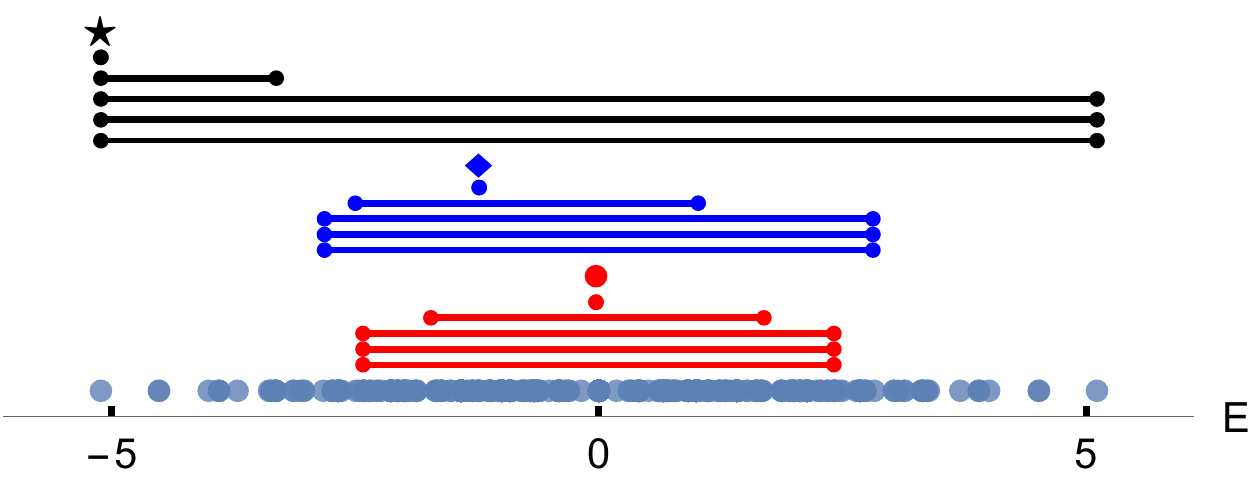}  \\ 
\hline
\end{tabular}
\caption{\label{tab:plots_app2} The continuation of Table~\ref{tab:plots_app}. 4) In the XY and PXP models, the local particle number measurement ($k=0$) does not determine the ground state energy well. This is because low and high-energy eigenstates produce similar, or the same in the case of PXP, distribution of outcomes.  5) Estimating the hot state energy in the PXP model works significantly better than estimating the ground state energy. This is because of the high degeneracy in the middle of the spectrum. 6) The ground state-optimized method performs better for ground states than the observable-optimized method but performs worse on average.}
\end{table*}

\ssect{a. Ground state-optimized measurements.}
First, we introduce a method that is $k$-local optimization for a specific state, in our case, the ground state. This method is inspired by the Matrix Product State ansatz~\cite{orus2014a}, and by the correspondence between observational and entanglement entropy~\cite{schindler2020quantum}. 

The logic of the motivation goes as follows: low observational entropy means that the system state wandered into one of the small subspaces-macrostates given by the measurement~\cite{safranek2019b,safranek2021generalized}. This means that we can estimate the maximal and the minimal value of the estimated observable in that subspace, which, in turn, translates into estimating these bounds for the system state itself. In other words, lower observational entropy means better estimates. According to Ref.~\cite{schindler2020quantum}, observational entropy minimized over local coarse-grainings leads to entanglement entropy, and the minimum is achieved when the local coarse-grainings are given by the Schmidt basis. This means that the measuring in the Schmidt basis will yield small observational entropy and, in turn, a better estimate of the mean value of observable. Thus, we need to find $k$-local measurements that reflect the Schmidt basis, which are expected to perform well in the estimation.

We do this as follows: Assuming we have a chain of length $L$ (for example, $L=10$) and some divisor $k\leq 10$ (for example $k=2$), we first divide the chain into two parts: one --- system $A_1$ ---  of length $k$ (i.e., sites $(1,2)$) and the other --- system $B_1$  --- of length $L-k$ (i.e., sites $(3,4,5,6,7,8,9,10)$). We define $\R_0$ as the ground state (or any other state we want to optimize for). We compute the reduced density matrix
\[
\R_{A_1}=\tr_{B_1}\R_0,
\]
and diagonalize it. The eigenbasis of the reduced density matrix is, by definition, the system $A_1$-local part of the Schmidt basis. We denote this basis as $\{\ket{\psi_{i_1}^{A_1}}\}$.

Then we move on to the next $k$ sites. We divide the system into $A_2$ (sites $(3,4)$) and $B_2$ (sites $(1,2,5,6,7,8,9,10)$). Again, we compute the reduced density matrix 
\[
\R_{A_2}=\tr_{B_2}\R_0,
\]
and find its eigenvectors, which we denote $\{\ket{\psi_{i_2}^{A_2}}\}$.

We continue dividing the system until the end (in our example, we go up to $A_5$). The final, ground state-optimized $k$-local measurement basis is then given by
\[
\{\ket{\psi_{i_1}^{A_1}}\otimes\ket{\psi_{i_2}^{A_2}}\otimes\ket{\psi_{i_3}^{A_3}}\otimes\ket{\psi_{i_4}^{A_4}}\otimes\ket{\psi_{i_5}^{A_5}}\}.
\]

\ssect{b. Observable-optimized measurements, type 1.}
Second, we introduce a method of $k$-local optimization for a specific observable, in our case, the Hamiltonian.

The motivation behind this optimization is that the estimation of the mean value of the observable works better the more the measurement resembles the eigenbasis of the observable. Thus, we create a procedure that generates a measurement that somewhat resembles the estimated observable.

We illustrate this method on the Heisenberg model, assuming $L=10$, which is given by the Hamiltonian (assuming hard-wall boundary conditions)
\[
\ham = \sum_{i=1}^9 \big(\hat \sigma_i^x\hat \sigma_{i+1}^x+\hat \sigma_i^y\hat \sigma_{i+1}^y+\hat \sigma_i^z\hat \sigma_{i+1}^z\big)+\sum_{i=1}^{10} h_i\hat \sigma_i^z.
\]

For $k=1$-local measurement, we remove all the terms spanning more than a single site. This leads to a modified Hamiltonian
\[
\ham_1 = \sum_{i=1}^{10} h_i\hat{\sigma}_i^z.
\]
We call the eigenbasis of this Hamiltonian the $k=1$-local observable optimized measurement for the Hamiltonian. Incidentally, in this case, this basis is precisely the same as the computational basis.

For $k=2$-local measurement, we divide the lattice into blocks of two sites and remove all the terms that cross those blocks. This leads to
\[
\ham_2 = \sum_{i=1,3,5,7,9} \big(\hat \sigma_i^x\hat \sigma_{i+1}^x+\hat \sigma_i^y\hat \sigma_{i+1}^y+\hat \sigma_i^z\hat \sigma_{i+1}^z\big)+\sum_{i=1}^{10} h_i\hat \sigma_i^z.
\]
The eigenbasis of this Hamiltonian is the $k=2$-local observable optimized measurement for the Hamiltonian.

For $k=5$-local measurement, we divide the lattice into two blocks of five sites and remove all the terms that cross those blocks. This leads to
\[
\ham_5 = \sum_{i=1,\dots,4,6,\dots,9} \big(\hat \sigma_i^x\hat \sigma_{i+1}^x+\hat \sigma_i^y\hat \sigma_{i+1}^y+\hat \sigma_i^z\hat \sigma_{i+1}^z\big)+\sum_{i=1}^{10} h_i\hat \sigma_i^z.
\]
The eigenbasis of this Hamiltonian is the $k=5$-local observable optimized measurement for the Hamiltonian.

In the case of $k=10$-local measurement, the corresponding Hamiltonian is the original Hamiltonian itself,
\[
\ham_{10}=\ham.
\]
Measuring in the basis of this Hamiltonian is the same as measuring the Hamiltonian itself, which yields perfect precision in estimating its mean value.

\begin{figure}[t!]
\begin{tabular}{|c|c|}
\hline
\multicolumn{2}{|c|}{initial states: \quad  \color{black}{$\filledstar$} G \quad  {\color{blue}{$\filleddiamond$}} C 
 \quad {\color{red}{$\bullet$}} H}\\
\hline
\multirow{2}{*}{Ham.} & large system \\
\cline{2-2}
 & observable-optimized (type 2) \\
\hline
 \begin{turn}{90}
PXP
\end{turn}   & \includegraphics[width=.8\hsize]{{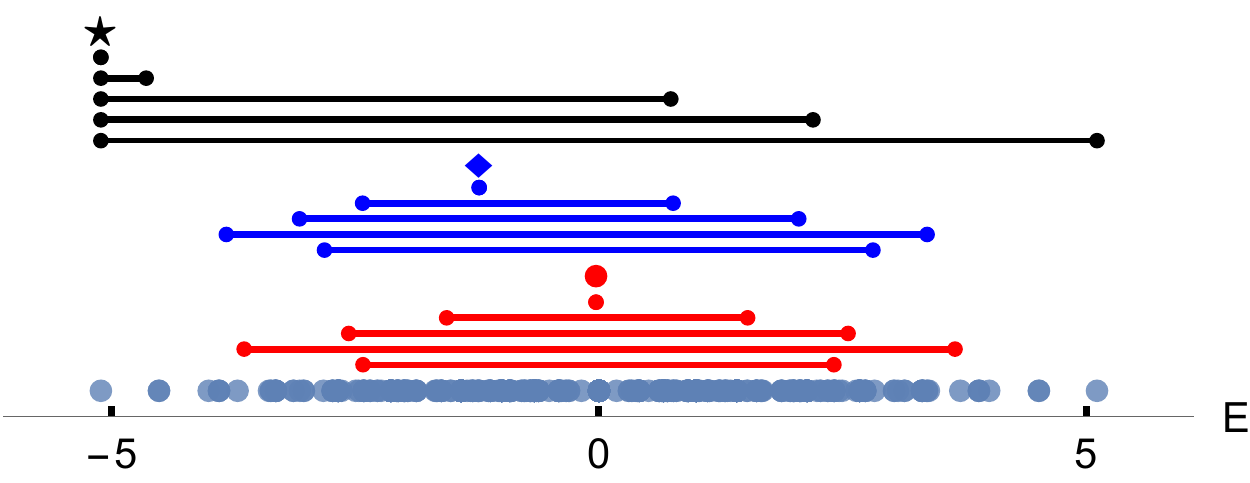}} \\
\hline
\end{tabular}
\caption
{Observable-optimized type 2 method for the large system of the PXP model. The PXP model is the only of our considered models in which type 1 and type 2 differ. Notice the worse performance for $k=1$ compared to $k=0$ for the cold and hot states and better performance for $k=1,2$ compared to the type 1 method for the ground state.}
\label{Fig:PXPtype2}
\end{figure}

\ssect{c. Observable-optimized measurements, type 2.}
Alternatively, one can consider a different way of finding observable-optimized measurements, somewhat similar to the ground-state optimization method. For $k=2$ and $L=10$, divide the system into system $A_1$ (the first two sites $(1,2)$) and system $B_1$ (the last $L-k$ sites $(3,4,5,6,7,8,9,10)$). Then compute the ``reduced Hamiltonian''
\[
\ham^{A_1}=\tr_{B_1}\ham,
\]
and diagonalize it, obtaining its eigenbasis $\{\ket{\psi_{i_1}^{A_1}}\}$. Continue analogously as in the ground-state optimization method to generate the global observable-optimized basis for $k=2$,
\[
\{\ket{\psi_{i_1}^{A_1}}\otimes\ket{\psi_{i_2}^{A_2}}\otimes\ket{\psi_{i_3}^{A_3}}\otimes\ket{\psi_{i_4}^{A_4}}\otimes\ket{\psi_{i_5}^{A_5}}\}.
\]

In our numerical experiments, due to the specific forms of the Hamiltonian, type 1 and type 2 observable-optimized measurements differ only in the PXP model. See Fig.~\ref{Fig:PXPtype2}. 

\ssect{d. Generating the unitary.}
Finally, we want to transform the optimized measurement and express it as a combination of a unitary operator applied to the system's state and, after that measuring in the computational basis. This is illustrated in Fig.~\ref{Fig:allowed_op}. Assuming our example $k=2$ and $L=10$ again, we start deriving the formula for $U^{A_1}$ with requiring that the probability of an outcome is the same in both situations:
\[
\bra{j_1,j_2}U^{A_1}\R U^{A_1\dag}\ket{j_1,j_2}\overset{!}{=}\bra{\psi_{i_1}^{A_1}}\R \ket{\psi_{i_1}^{A_1}},
\]
where $\ket{j_1,j_2}$ is the computational basis vector, $\ket{\psi_{i_1}^{A_1}}$ an optimized basis vector on the first two sites, and we want to match each couple $j_1,j_2$ to one $i_1$. A sufficient condition is
\[
\ket{\psi_{i_1}^{A_1}}=U^{A_1\dag}\ket{j_1,j_2}.
\]
Assuming that $\ket{\psi_{i_1}^{A_1}}$ is a column vector written in the computational basis and that each site is a qubit (which leads to $i_1=1,2,3,4$), we have
\[
U^{A_1\dag}=
\begin{pmatrix}
 & & & \\
    \ket{\psi_{1}^{A_1}} & \ket{\psi_{2}^{A_1}} & \ket{\psi_{3}^{A_1}} & \ket{\psi_{4}^{A_1}} \\
 & & & \\
\end{pmatrix}.
\]
This means that
\[
U^{A_1}=
\begin{pmatrix}
\ket{\psi_{1}^{A_1}}^\dag\\
\ket{\psi_{2}^{A_1}}^\dag\\
\ket{\psi_{3}^{A_1}}^\dag\\
\ket{\psi_{4}^{A_1}}^\dag
\end{pmatrix}
=
\begin{pmatrix}
\bra{\psi_{1}^{A_1}}\\
\bra{\psi_{2}^{A_1}}\\
\bra{\psi_{3}^{A_1}}\\
\bra{\psi_{4}^{A_1}}
\end{pmatrix}.
\]
The global optimized measurement then consists of applying the unitary operation
\[
U=U^{A_1}\otimes \cdots \otimes U^{A_5}=\begin{pmatrix}
\bra{\psi_{1}^{A_1}}\\
\bra{\psi_{2}^{A_1}}\\
\bra{\psi_{3}^{A_1}}\\
\bra{\psi_{4}^{A_1}}
\end{pmatrix}\otimes \cdots \otimes 
\begin{pmatrix}
\bra{\psi_{1}^{A_5}}\\
\bra{\psi_{2}^{A_5}}\\
\bra{\psi_{3}^{A_5}}\\
\bra{\psi_{4}^{A_5}}
\end{pmatrix}
\]
on the system and then measuring in the computational basis.

\bibliography{main.bib}

\end{document}